\documentclass[twocol]{ametsocV6.1}
\usepackage{graphicx}
\usepackage{float}
\usepackage{booktabs}
\usepackage{hyperref}

\title{An analytic formula for entraining CAPE in mid-latitude storm environments}

\authors{John M. Peters\aff{a}\correspondingauthor{John M. Peters, John.M.Peters@psu.edu},  Daniel R. Chavas\aff{b}, Chun-Yian Su\aff{a}, Hugh Morrison\aff{c}, and Brice E. Coffer\aff{d}}

\affiliation{\aff{a}Department of Meteorology and Atmospheric Science, The Pennsylvania State University, University Park, PA \\ \aff{b} Department of Earth, Atmospheric, and Planetary Sciences, Purdue University, West Lafayette, IN \\
\aff{c} National Center for Atmospheric Research, Boulder, CO \\
\aff{d}Department of Marine, Earth and Atmospheric Sciences, North Carolina State University, Raleigh, NC }

\abstract{This article introduces an analytic formula for entraining convective available potential energy (ECAPE) with an entrainment rate that is determined directly from the storm environment.  Extending previous formulas derived in \citet{PMNMT2020}, entrainment is connected to the background environment via an analytic manipulation of the equations of motion that yields a direct correspondence between the storm relative flow and the updraft radius, and an inverse scaling between the updraft radius squared and entrainment rate. These concepts, combined with the assumption of adiabatic conservation of moist static energy, yield an explicit analytic equation for ECAPE that depends entirely on state variables in an atmospheric profile and a few constant parameters with values that are established in past literature.  Using a simplified Bernoulli-like equation, a second formula is derived that accounts for updraft enhancement via kinetic energy extracted from the cloud's background environment. CAPE and ECAPE can be viewed as predictors of the maximum vertical velocity $w_{max}$ in an updraft.  Hence, these formulas are evaluated using $w_{max}$ from past numerical modeling studies.  Both of the new formulas improve predictions of $w_{max}$ substantially over undiluted CAPE, ECAPE with a prescribed entrainment rate, and the ECAPE formula from \citet{PMNMT2020}.  The formula that incorporates environmental kinetic energy contribution to the updraft correctly predicts instances of exceedance of $\sqrt{2\text{CAPE}}$ by $w_{max}$ in simulations, and provides a conceptual explanation for why such exceedance is rare among past simulations.  These formulas are potentially useful in nowcasting and forecasting thunderstorms and as thunderstorm proxies in climate change studies.}

\begin{document}

%% Necessary!
\maketitle

%%%%%%%%%%%%%%%%%%%%%%%%%%%%%%%%%%%%%%%%%%%%%%%%%%%%%%%%%%%%%%%%%%%%%
% SIGNIFICANCE STATEMENT/CAPSULE SUMMARY
%%%%%%%%%%%%%%%%%%%%%%%%%%%%%%%%%%%%%%%%%%%%%%%%%%%%%%%%%%%%%%%%%%%%%
%
% If you are including an optional significance statement for a journal article or a required capsule summary for BAMS 
% (see www.ametsoc.org/ams/index.cfm/publications/authors/journal-and-bams-authors/formatting-and-manuscript-components for details), 
% please apply the necessary command as shown below:
%
% Significance Statement (all journals except BAMS)
%
\statement
Substantial mixing occurs between the upward moving air currents in thunderstorms (updrafts) and the surrounding comparatively dry environmental air, through a process called entrainment.  Entrainment controls thunderstorm intensity via its diluting effect on the buoyancy of air within updrafts.  A challenge to representing entrainment in forecasting and predictions of the intensity of updrafts in future climates is to determine how much entrainment will occur in a given thunderstorm environment without a computationally expensive high resolution simulation.  To address this gap, this article derives a new formula that computes entrainment from the properties of an updraft's background environment.  This formula is shown to predict updraft vertical velocity more accurately than past diagnostics, and can be used in forecasting and climate prediction to improve predictions of thunderstorm behavior and impacts. 
%	 Enter significance statement here, no more than 120 words. See \url{www.ametsoc.org/index.cfm/ams/publications/author-information/significance-statements/} for details.
%
%% Capsule (BAMS only)
%%
%\capsule
%       Enter BAMS capsule here, no more than 30 words. See \url{www.ametsoc.org/index.cfm/ams/publications/author-information/formatting-and-manuscript-components/#capsule} for details.
% x
%% * * If using twocol mode, you will need to use the commands "twocolsig" and "twocolcapsule" in place of "sig" and "capsule"
%%      to ensure that the text box correctly spans across both columns.
%

%%%%%%%%%%%%%%%%%%%%%%%%%%%%%%%%%%%%%%%%%%%%%%%%%%%%%%%%%%%%%%%%%%%%%
% MAIN BODY OF PAPER
%%%%%%%%%%%%%%%%%%%%%%%%%%%%%%%%%%%%%%%%%%%%%%%%%%%%%%%%%%%%%%%%%%%%%
%

\section{Introduction}

Middle-to-upper\footnote{We contrast middle-to-upper tropospheric vertical velocities, which are primarily buoyantly driven, with lower tropospheric vertical velocities which are often dynamically driven in squall lines \citep[e.g.,][]{BR2014,JR2015} and supercells \citep[e.g.,][]{WR2000,PNM2019}.} tropospheric vertical velocities in deep convective updrafts influence a variety of storm-related societal impacts, including precipitation \citep[e.g.,][]{ET2022}, hail \citep[e.g.,][]{Danielsen:1972,LK2022}, electrification \citep[e.g.,][]{RSVM2014,STOLZ2015}, downdraft and cold pool intensity \citep[e.g.,][]{MT2019}, tropospheric convective mass flux \citep[e.g.,][]{PMZ2020}, and the flux of mass, aerosols, and water vapor across the tropopause \citep[e.g.,][]{MHJLT2013}.  The magnitude of vertical velocities in the upper reaches of deep convective updrafts are strongly influenced by updraft buoyancy \citep[e.g.,][]{MP2017,PNM2019,J2017}.  It is well known that entrainment-driven dilution of deep convective updrafts substantially influences updraft buoyancy and vertical velocity \citep[e.g.,][]{Zipser2003,RC2010,RK2010nn}.  For instance, weakly sheared deep convective updrafts with large fractional entrainment rates are substantially diluted and often only realize a small fraction (e.g., 20-30 \%) of their convective available potential energy (CAPE) as updraft kinetic energy KE \citep[][]{RC2010}. In contrast, more organized modes of deep convection such as squall lines and supercells with smaller fractional entrainment rates and less dilution can realize much larger fractions of their CAPE as KE \citep[i.e., 80-100 \%][]{LM2015,PNM2019,MPM2021}.  Hence, storm-to-storm variations in entrainment substantially alter how much CAPE a storm is able to process, and consequently its updraft kinetic energy and vertical velocity.  These storm-to-storm variations in entrainment also generally supersede the influences of variations in other updraft processes and environment factors on vertical velocity that receive substantial attention in the literature \citep[e.g.,][]{LEB2018,GM2021}, such as aerosol effects, pressure perturbations, and precipitation behavior.  Hence, the atmospheric sciences community would benefit from an accurate representation of entrainment in research and forecasting diagnostic parameters, such as CAPE, so that the parameters can more accurately characterize the intensity of convective updrafts that might form in a given environment. %This suggest that research and forecasting applications should use entraining CAPE (ECAPE), which is CAPE that is adjusted to account for the effects of entrainment on parcel buoyancy, in lieu of undiluted CAPE.

CAPE calculations that include entrainment effects are referred to as entraining CAPE, or ECAPE.  Whereas CAPE is often viewed as the theoretical maximum kinetic energy that can be extracted by an isolated parcel from its environment, ECAPE makes additional assumptions about updraft steadiness and mixing to estimate how the efficiency of this kinetic energy extraction is affected by entrainment. Various ECAPE-like calculations have been used for the better part of the last century, primarily in the climate, tropical meteorology, and cumulus parameterization communities.  For instance, simple plume models \citep[e.g.,][]{ST1962} for moist convective updrafts predict profiles of buoyancy that include entrainment effects, which can be vertically integrated to obtain ECAPE. The ``cloud work function'', which is an essential element of many cumulus parameterizations \citep{AS1974}, uses the buoyancy of a diluted parcel within its calculation, and yields a quantity that is analogous to ECAPE.  ECAPE is used as diagnostic tool in the research of tropical environments to explain the sensitivity of deep convection initiation to free tropospheric moisture \citep[][]{BZ1997}, and in the closure formulation of cumulus parameterizations \citep{ZHANG2009}. The zero-buoyancy plume model, in which buoyancy is assumed to be exactly extinguished by entrainment, yields analytic solutions for the mean state thermal structure of the tropical atmosphere \citep{SO2013}.  The range of fractional entrainment rates in the tropics is typically smaller than that of the mid latitudes \citep[e.g.,][]{TZG2021}.  Hence, using an ECAPE calculated with an empirically obtained constant fractional entrainment rate provides reasonably accurate predictions of deep convective updraft characteristics in the tropics \citep[e.g.,][]{G2001}

There are also a few scattered applications of ECAPE in the weather forecasting community. For instance, the spatial distribution of ECAPE has been shown to better identify the tornadic regions of tropical \citep[][]{SKNTYPHOON} and extratropical cyclones \citep[][]{EKNExtratropical} than undiluted CAPE.  ECAPE has also been used to predict vertical velocities in supercells more accurately than standard CAPE calculations \citep{PMNMT2020}.  There is substantially larger variability in fractional entrainment in the continental mid-latitudes \citep[e.g.,][]{PNM2020,TZG2021,TEABT2021} than in the tropics, meaning that ECAPE computed with a single fractional entrainment rate cannot accurately describe all midliatude convective environments \citep[e.g.,][]{PNM2020}.  This makes using ECAPE in midlatudes more difficult than in the tropics, because it is not always clear what entrainment rate should be used in the calculation.

To address the issue over what choice of fractional entrainment rate to use in the midlatitudes, \citet{PMNMT2020} (hereafter P20) developed an analytic formula for maximum updraft vertical velocity (which is equal to $\sqrt{2 ECAPE}$) that calculated entrainment from attributes of a storm's background environment, rather than requiring that the user specify an entrainment rate. The connection between entrainment and the background environment in this formula was based on the previously-established negative correspondence between vertical wind shear and fractional entrainment \citep[e.g.,][]{PNM2019,PNM2020,PMNMMN21_1,PMNMMN21_2}.  That is, mature deep convective updrafts tend to be wider in environments with strong vertical wind shear and have accordingly smaller fractional entrainment rates.  This formula more accurately predicted maximum updraft vertical velocities than standard ECAPE computed with constant pre-specified fractional entrainment rate.

There are several shortcomings of the P20 study that warrant a revisit of the concepts contained therein.  First, the expression derived in the paper uses a hodgepodge of formulas from previous studies, such as \citet{MOR2016c} and \citet{PNM2019} as a starting point\footnote{Note a litany of constants are carried over into P20 from these past formulas, and some of the symbols used (such as $H_v$ for the latent heat of vaporization) are inconsistent with the symbols used in some of our more recent articles \citep[e.g., $L_v$ for the latent heat of vaporization][]{PC2021,PCM2021,PMNMMN21_1}).}.  The assumptions underlying these formulas from previous studies are not explicitly discussed in P20, nor are they even thoroughly scrutinized in their source articles.  Because of this rooting in past studies, a few of the terms that end up in the P20 equation are complicated and lack obvious physical underpinning, which is challenging for end users of this formula.   

Second, the end formula for maximum updraft vertical velocity is a third-order polynomial equation that must either be solved explicitly with the complicated quartic equation, or with a numerical root finding procedure.  End users of the formula found this quartic solution difficult to efficiently incorporate into software routines.  This 3rd order polynomial equation results from the assumption that fractional entrainment $\varepsilon$ scales with the inverse of updraft radius $R^{-1}$.  However, there is now evidence that $\varepsilon \sim R^{-2}$ is a more realistic scaling \citep[][]{PNM2019,MPCS2021,MPM2021}.  Re-formulating the P20 equation with $\varepsilon \sim R^{-2}$ yields a 2nd-order polynomial equation that is much easier to solve, as will be shown in the present study.

Third, the title of that paper, which is ``A formula for the maximum vertical velocity in supercell updrafts'', obscures the take-home messages.  The title does not contain the terms entrainment or CAPE, so it is not obvious that the parameter derived in the paper essentially modifies CAPE to account for the effects of entrainment (which is by definition ECAPE).  The concepts contained within the paper apply to any isolated deep convective updraft existing within moderate to strong vertical wind shear -- they are not limited to supercells.  There is no assumption about updraft rotation within the mathematical framework.  Hence, the inclusion of the term supercell in the title made the application of the formula sound unnecessarily restrictive.  

Our goal in this article is to revisit the concepts of P20 to derive ECAPE formulas (Sections 2-3) that improve upon the concepts in the P20 study in the following ways:
\begin{enumerate}
    \item 
    The buoyancy formula in the present study is derived directly from the assumed conservation of moist static energy, which differs from the P20 formula which used the supersaturation tendency equation from \citet{PC1988} as a starting point.  This methodological alteration requires less severe assumptions and results in formulas with greater accuracy in the present study.
    \item
    The new formula uses the $\varepsilon \sim R^{-2}$ scaling, with further improves accuracy over the P20 formula.
    \item
    We also account for additional processes that were not considered by P20, such as the contribution to updraft kinetic energy from the kinetic energy an updraft extracts from its inflow via pressure gradient accelerations. 
\end{enumerate}
The new ECAPE formulas are evaluated with output from four past numerical modeling studies that included 141 simulations (Section 4).  The formulas and their constituent terms, along with recommended parameter values, are summarized in the discussion and conclusions (Section 5).    

\section{Derivation of analytic ECAPE formula}

The derivation relies on three underlying concepts:  a scaling between entrainment and updraft radius (section 2a), an analytic relationship between ECAPE and entrainment (section 2b), and an analytic relationship between updraft radius and state variables within an atmospheric sounding (sections 2c-d).  Combining these components allows us to eliminate entrainment and updraft radius to express ECAPE as a function of the state variables within a sounding.

%\subsection{Benchmarks for evaluating our derived formulas}

We will need to make numerous approximations through the course of the derivation.  To evaluate the accuracy of these approximations, we will first establish a benchmark calculation of both buoyancy and ECAPE computed with as few approximations as possible.  This benchmark calculation uses the adiabatic unsaturated and saturated lapse rate equations derived in \citet{PCM2021}, eqs. 19 and 24 from that article respectively, with a mixed-phase layer in the parcel temperature range of 273.5 K to 233.15 K (see that study for details on the mixed-phase calculation), and the bulk plume entrainment approximation for the mixing of individual state variables with that of a horizontally invariant background environment (see eq. 36-38 in that study).  %The mathematical framework underpinning these formulas uses the fewest approximations possible in a parcel theory calculation \citep[][]{RKAPPROX}.

The formulas are evaluated using the severe weather proximity sounding dataset of \citet{THOM2003}.  This dataset includes 1028 atmospheric profiles taken near severe weather events that ranged from disorganized deep convection to tornadic supercells.  In each profile, the parcel with the largest undiluted CAPE in lowest 5 km of the atmosphere is lifted to calculate buoyancy, CAPE, and ECAPE.

\subsection{Connecting fractional entrainment to updraft radius}

Our first step is to establish a relationship between updraft radius and the fractional entrainment rate $\varepsilon$. We accomplish this by deriving an expression for passive tracer dilution in the cloud core assuming that entrained air has a tracer value of zero, and assuming that detrained air has a tracer value equal to that locally in the cloud core. Here $\varepsilon$ is the fractional entrainment rate needed to produce a vertical profile of cloud core passive tracer consistent with the dilution it undergoes. %We emphasize that such bulk calculations of $\varepsilon$ (``bulk'' meaning that all entrained air at a given height has characteristics of the background environment at that height) differ from $\varepsilon$ calculated directly by the mass flux into a cloud \citep{ROMPS2010}. Although bulk estimates of $\varepsilon$ are generally smaller than direct calculations by about a factor of 2, the shape of bulk and direct entrainment profiles are similar \citep{ROMPS2010}. Moreover, we are ultimately concerned with cloud core dilution rather than the mass flux into the cloud, and the bulk calculation of $\varepsilon$ is self-consistent with the dilution of cloud core tracer given a specified environmental tracer profile.

The derivation closely follows that of \citet{MOR2016c} (hereafter M17), section 2a therein.  We first consider a passive tracer $C$, whose mixing ratio (in kg kg\textsuperscript{-1}) is 1 in a cloud's effective inflow layer \citep[i.e., the layer of nonzero CAPE][]{TCR2007,NPM2020}, and 0 above this layer.  Conceptually, the passive tracer value represents the degree to which a parcel has been diluted via entrainment, with $C\approx 1$ indicating undiluted air, and $C<<1$ indicating highly diluted air.

The anelastic Lagrangian tendency equation for $C$ may be written in cylindrical coordinates as:
\begin{equation}
\label{eq:PTRA1}
\frac{dC}{dt} =\frac{\partial C}{\partial t} + \frac{1}{r}\frac{\partial r u C}{\partial r} + \frac{1}{r} \frac{\partial vC}{\partial \phi}+ \frac{1}{\rho_0}\frac{\partial \rho_0 w C}{\partial z} = 0, 
\end{equation}
where $r$, $\phi$, and $z$ are the radial, azimuthal, and vertical coordinates, $u$, $v$, and $w$ are the corresponding radial, azimuthal, and vertical velocities, and $\rho_0(z)$ is a reference density profile.  Azimuthally averaging this equation, and then Reynolds averaging, yields:
\begin{equation}
\label{eq:PTRA1}
\frac{d \overline{C}}{dt} = -\frac{1}{r}\frac{\partial r\overline{u'C'}}{\partial r} - \frac{1}{\rho_0}\frac{\partial \overline{\rho_0 w'C'}}{\partial z} 
\end{equation}
where overbar denotes a spatial average with a filter scale similar to that of the updraft width (tyically on the order of 1-2 km), primes denote deviations smaller than the filter scale, and $\frac{d\overline{C}}{dt} = \frac{\partial \overline{C}}{\partial t} + \overline{u}\frac{\partial \overline{C}}{\partial r} + \overline{v}\frac{\partial \overline{C}}{\partial \phi} + \overline{w}\frac{\partial \overline{C}}{\partial z}$.  Physically, the overbar terms correspond to updraft-scale flow patterns, whereas the $'$ terms correspond to turbulent fluxes.  We neglect the vertical turbulent flux term since recent large eddy simulations have supported a dominant role of lateral mixing in entrainment \citep[][]{BJNS2014}.  All quantities are valid at the updraft horizontal center unless explicitly stated otherwise.

Following M17 and \citet{DE2010}, we assume that $\overline{u'C'}$ varies linearly over a turbulent mixing length scale $L_{mix}$ and vanishes at the updraft center, such that $\overline{u'C'}(r) = \left.\overline{u'C'}\right|_{L_{mix}} \left(\frac{r}{L_{mix}}\right)$, where the $\left.\overline{u'C'}\right|_{L_{mix}}$ denotes the value of $\overline{u'C'}$ at distance $L_{mix}$ from the updraft center.  Finally, we use the chain rule to write $\frac{\overline{d}}{dt} = \overline{w} \frac{\overline{d}}{dz}$, where $\frac{\overline{d}}{dz}$ is the rate of change of a quantity as the parcel changes height.  Making these approximations allows us to write eq. \ref{eq:PTRA1} as:
\begin{equation}
\label{eq:PTRA2}
\frac{d \overline{C}}{dz} =  -2  \frac{\left.\overline{u'C'}\right|_{L_{mix}}}{\overline{w} L_{mix}}. 
\end{equation}
In the eddy diffusivity approximation \citep[e.g.,][]{K1962}, we assume that turbulent fluxes act to diffuse a quantity down-gradient.  Using this approach, we may write $\left.\overline{u'C'}\right|_{L_{mix}} \approx -\frac{k^2 L_{mix}^2}{P_r} \left| \frac{\partial w}{\partial r}\right| \frac{\partial C}{\partial r}$ (eqs. 5-6 in M17) and eq. \ref{eq:PTRA2} as:
\begin{equation}
\label{eq:PTRA3}
\frac{d \overline{C}}{dz} = 2 \frac{k^2 {L_{mix}}}{\overline{w} P_r} \left| \frac{\partial w}{\partial r}\right| \frac{\partial C}{\partial r}, 
\end{equation}
where $k^2$ is the von Karman constant and $P_r$ is the turbulent Prandtl number. Finally, we use linear approximations to the lateral gradients in $C$ and $w$, such that $\frac{\partial C}{\partial r} = \frac{C_0 - \overline{C}}{R}$ and $\left|\frac{\partial w}{\partial r}\right| = \frac{|w_0-\overline{w}|}{R}$, and assume that $w_0=0$ and $C_0 = 0$ to write:
\begin{equation}
\label{eq:PTRA3}
\frac{d\overline{C}}{dz} = - \varepsilon \overline{C},
\end{equation}
where
\begin{equation}
\label{eq:Avarepsilon}
\varepsilon = \frac{2 k^2 L_{mix}}{P_r R^2}.
\end{equation}
Equation \ref{eq:PTRA3} takes the form of a classical steady-state plume equation \citep[][]{ST1962,B1975}, where $\varepsilon$ is the fractional entrainment inverse length scale.  This term represents the rate at which $C$ is diluted with height by entrainment.  There is some debate in past literature over how $L_{mix}$ should be interpreted.  For instance, in \citet{CHAINPT1_2019}, P20, and \citet{PMZ2020}, we simply set $L_{mix} \sim R$, which from Equation \ref{eq:Avarepsilon} results in a $\varepsilon \sim R^{-1}$ scaling.  However, analysis of large eddy simulations (LES) in our more recent work \citep[e.g.,][]{MPM2021,MPCS2021} indicates that $\varepsilon \sim R^{-2}$, suggesting from Equation \ref{eq:Avarepsilon} that $L_{mix}$ should be viewed as a constant.  Hence, we set $L_{mix}$ to a fixed value following \citet{MPCS2021}.  

The eddy diffusivity approximation for lateral mixing implicitly neglects the entrainment of air occurring within organized updraft-scale flow, which is known as dynamic entrainment \citep[e.g.,][]{ROOY2013}.  However, our past work has shown that dynamic entrainment primarily affects updraft properties below the height of maximum $w$ where flow is laterally convergent into the updraft \citep[e.g.,][]{MOR2016c,CHAINPT1_2019,MPCS2021}.  Hence, it is reasonable to neglect dynamic entrainment in our present objective of deriving an expression for ECAPE, which pertains to the maximum kinetic energy achieved by the updraft that coincides with the position of maximum $w$.

\subsection{Derivation of analytic expressions for the buoyancy and $ECAPE$ of an entraining parcel}

Our next step is to express ECAPE as an analytic function of $\varepsilon$, wherein $\varepsilon$ is not contained within integrals or differentials.  We begin with the first law of thermodynamics for a rising parcel, which may be written as \citep[e.g.,][]{EMAN94,ROMPS2015,PCM2021}:
\begin{equation}
\label{eq:firstlaw}
c_{pm} \frac{dT}{dz} - \frac{1}{\rho} \frac{dp}{dz} + L_v\frac{dq_v}{dz} - L_i \frac{dq_i}{dz} = Q 
\end{equation}
where $c_{pm}$ is the moist heat capacity that depends on water vapor and condensates, $T$ is temperature, $\rho$ is density, $p$ is pressure, $L_v$ is the temperature dependent latent heat of vaporization, $q_v$ is the water vapor mass fraction, $L_i$ is the temperature dependent latent heat of freezing, $q_i$ is the ice mass fraction, $Q$ represents all diabatic effects, and $\frac{d}{dz}$ represents the rate at which a quantity changes as a parcel changes its vertical position.

We simplify this equation by making a series of approximations.  First, we replace the moist heat capacity $c_{pm}$ with the constant dry-air heat capacity $c_{pd}$.  Second, we use the hydrostatic equation to write $\frac{1}{\rho} \frac{dp}{dz} = -g$, where $g$ is the acceleration of gravity.  Third, we neglect ice ($q_i=0$).  Fourth, we replace the temperature-dependent latent heat of vaporization with its reference value at the triple point temperature $L_{v,r}$.  Fifth, we assume that the only diabatic effect is the mixing of a parcel with its far-field environmental profile.  Using these approximations, we may re-write eq. \ref{eq:firstlaw} as:
\begin{equation}
\label{eq:mseten}
\frac{dh}{dz} = - \varepsilon \left( h - h_0 \right),
\end{equation}
where $h$ is the moist static energy, defined as
\begin{equation}
\label{eq:mseapprox}
    h = c_{pd} T + L_{v,r} q + gz,
\end{equation}
$h_0$ is the moist static energy of the background environment, defined as:
\begin{equation}
\label{eq:mseenv}
    h_0 = c_{pd} T_0 + L_{v,r} q_0 + gz,
\end{equation}
the subscripts $0$ denote the height-dependent background environmental profile, and we have dropped the $v$ subscript on $q$ for simplicity.  The $- \varepsilon \left( h - h_0 \right)$ term represents dilution of $h$ with height due to entrainment, and is expressed in a manner consistent with a classical plume updraft model \cite[e.g.,][]{B1975}.  Note that for an adiabatic parcel (i.e., $\varepsilon \rightarrow 0$), $h$ is conserved.  Hence, $h$ is analogous to equivalent potential temperature ($\theta_e$). It will also be useful later to define the saturated moist static energy of the environment $h_0^*$ as:
\begin{equation}
\label{eq:mseenvsat}
    h_0^* = c_{pd} T_0 + L_{v,r} q_0^* + gz,
\end{equation}
where $q^*$ is the saturation mass fraction defined via eq. 10 in \citet{BOLT1980}.  Finally, we define the buoyancy $B$ of an updraft air parcel as:
\begin{equation}
\label{eq:buoyancy}
B = g \frac{T - T_0}{T_0},
\end{equation}
which neglects the effects of water vapor and condensate loading on buoyancy.

\begin{figure*}[!ht]
\centerline{\includegraphics[width=40pc]{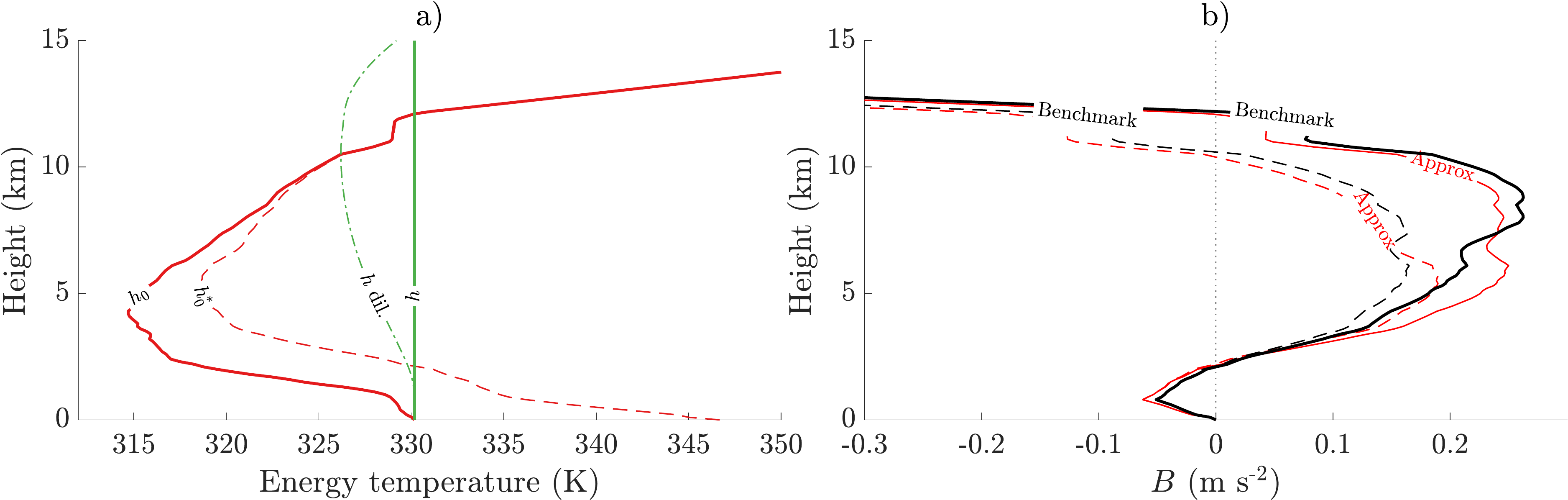}}
\caption{Panel a: profiles of environmental $h_0$, $h_0^*$, and $h$ of an undiluted parcel, and the $h$ of a diluted parcel with $\varepsilon = 1\times 10^{-4}$ m\textsuperscript{-1} (``h dil.''), computed using the tornadic supercell composite profile from \citet{P2014vortex}.  Moist static energies have been divided by $c_{pd}$ to yield ``energy temperature" with units of K.  Panel b: buoyancy of the diluted (dashed lines) and undiluted (solid lines) parcels, computed using the benchmark parcel (black, described in the beginning of this section) and from the approximate formula for $h$ calculated by numerically integrating eq.\ref{eq:mseten} as described in the text (red).} \label{parcel_path}
\end{figure*}

To evaluate the accuracy of these approximate equations, we integrate eq. \ref{eq:mseten} upward using a forward Euler integration scheme with a vertical grid spacing of 100 m, and solve for $T$ at each height using a numerical nonlinear equation solver.  We use $\frac{dq}{dz} = - \varepsilon \left(q - q_0 \right)$ during the unsaturated part of parcel ascent, and set $q=q^*$ during the saturated part of parcel ascent.   Quantities such as buoyancy and ECAPE computed with \ref{eq:mseten} and eq. \ref{eq:buoyancy} are referred to as ``approximate''.  The vertical distributions of $h_0$ and $h_0^*$ in a typical deep convective environment are shown in Fig. \ref{parcel_path}a.  Much like the typical vertical distribution of $\theta_e$, $h$ has a local maximum in the lower troposphere when nonzero CAPE is present, a local minimum in the middle troposphere, and becomes large again in the lower stratosphere.  An undiluted parcel lifted from the surface has larger $h$ than its surroundings until it reaches the lower stratosphere.  In an entraining parcel, $h$ gradually relaxes to that of the background environment as the parcel ascends.   Profiles of approximate buoyancy are compared to benchmark buoyancy, calculated from equations in \citet{PCM2021} as described earlier in this section, for undiluted and diluted parcels in Fig. \ref{parcel_path}b.  Despite the assumptions made thus far, the approximate and benchmark buoyancy profiles are comparable, having similar profile shapes and magnitudes at all heights. 

Combining eqs. \ref{eq:mseapprox}, \ref{eq:mseenv}, and eq. \ref{eq:mseenvsat} yields:
\begin{equation}
\label{eq:bdefsat}
B = \frac{g}{c_{pd} T_0} \left(h - h_0^* \right) -  \frac{g L_{v,r}}{c_{pd} T_0} \left( q^*  - q_0^*\right),
\end{equation}
where we have assumed that the updraft parcel is saturated, such that $q=q^*$.  The second term on the RHS of eq. \ref{eq:bdefsat} is often small relative to the first.  Hence, eq. \ref{eq:bdefsat} suggests that $B>0$ when $h > h_0^*$.  This agrees with Fig. \ref{parcel_path}a-b, which shows approximate coincidence between the vertical extent of $h > h_0^*$ (Fig. \ref{parcel_path}a) and the vertical extent of $B>0$ (Fig. \ref{parcel_path}b).  An entrainment term (i.e., $\varepsilon$) does not show up explicitly in eq. \ref{eq:bdefsat}, but is included implicitly via the moist static energy of the updraft parcel $h$, which is affected by entrainment.  To make $\varepsilon$ show up explicitly, we find the particular solution to eq. \ref{eq:mseten} with $h=h_0$ at $z=0$, which may be written as:
\begin{equation}
\label{eq:hanalytic}
h = e^{-\varepsilon z} \left(h_{ud} + \int_{\xi=0}^{\xi=z} \varepsilon e^{\varepsilon \xi} h_0 d\xi \right),
\end{equation}
where $h_{ud}$ is the moist static energy of an undiluted parcel (or equivalently the moist static energy of the entraining parcel at its origin height since we assume $h$ is conserved for undilute ascent),  $\xi$ is a dummy variable of integration, and we defined the parcel starting height as $z=0$ for simplicity.  Combining eq. \ref{eq:hanalytic} with eq. \ref{eq:bdefsat} yields the following:
\begin{equation}
\label{eq:banalytic_RR}
B = \frac{g}{c_{pd} T_0} \left[e^{-\varepsilon z} \left(  h_{ud} +  \int_{\xi=0}^{\xi=z} \varepsilon e^{\varepsilon \xi} h_0 d\xi \right) - h_0^* \right] -  \frac{g L_{v,r}}{c_{pd} T_0} \left( q^*  - q_0^*\right).
\end{equation}
The term $\varepsilon$ now shows up explicitly in the equation, but is contained within integrals.  We will need to make some additional approximations to bring this term out of the integrals to obtain our desired analytic solution.  

Eq. \ref{eq:banalytic_RR} can be re-arranged to express $B$ as a modification to the undiluted buoyancy $B_{ud}$ using eq. \ref{eq:bdefsat} evaluated with $h=h_{ud}$ and $q=q_{ud}$:
\begin{equation}
\label{eq:banalytic2}
\begin{aligned}
B = B_{ud} e^{-\varepsilon z}  + \frac{g}{c_{pd} T_0} \left( e^{-\varepsilon z} \int_{\xi=0}^{\xi=z} \varepsilon e^{\varepsilon \xi} h_0 d\xi  -  \left(1 - e^{-\varepsilon z}\right)h_0^* \right) \\
  -  \frac{g L_{v,r}}{c_{pd} T_0} \left( q^*  - q_0^*\right) +  e^{-\varepsilon z} \frac{g L_{v,r}}{c_{pd} T_0} \left( q_{ud}^*  - q_0^*\right).
  \end{aligned}
\end{equation}
This re-arrangement provides us with the opportunity to use the the undiluted buoyancy computed with the benchmark parcel to calculate $B_{ud}$ rather than the approximate $B_{ud}$ when evaluating eq. \ref{eq:banalytic2} (i.e., the black line in Fig. \ref{parcel_path} instead of the red line).  This substitution generally improves the accuracy of the formula, and is used in all subsequent calculations.

We note that the two terms on the RHS of eq. \ref{eq:banalytic2} will cancel each other in the limit of $\varepsilon \rightarrow 0$.  In the opposite limit of $\varepsilon \rightarrow \infty$, each of these terms individual vanish because $q^*\rightarrow q_0^*$ and $e^{-\varepsilon z}\rightarrow 0$.  We assume these terms are small in the intermediary range of $\varepsilon$, and consequently neglect them to simplify the equation.  Using integration by parts and neglecting the aforementioned terms, we may re-write eq. \ref{eq:banalytic2} as:
\begin{equation}
\label{eq:banalytic3}
\begin{aligned}
B = B_{ud} e^{-\varepsilon z}  + \frac{g}{c_{pd} T_0} \left(   \varepsilon z \widehat{h_0} +  e^{-\varepsilon z} \varepsilon^2 
  \int_{\xi=0}^{\xi = z} \widehat{h_0} \xi  e^{\varepsilon \xi} d \xi   -  \left(1 - e^{-\varepsilon z}\right)h_0^* \right) .
  \end{aligned}
\end{equation}
where $\widehat{h_0}(\xi) \equiv \frac{1}{\xi} \int_{\xi^*=0}^{\xi^*=\xi} h_0 d\xi^*$ is the average of $h_0$ below height $\xi$ and $\widehat{h_0}$ in the first term in the parentheses on the RHS is evaluated at $\xi = z$. If we assume that $\widehat{h_0}$ is approximately constant with height\footnote{This assumption is reasonable, given that vertical variations in $\widehat{h_0}$ are on the order of $1 \times 10^{4}$ J kg\textsuperscript{-1}, whereas the typical magnitude of this quantity is on the order of $1 \times 10^{6}$ J kg\textsuperscript{-1}.} in the integral term in eq. \ref{eq:banalytic3}, the equation simplifies dramatically to the following:
\begin{equation}
\label{eq:banalytic}
B = B_{ud} e^{-\varepsilon z}  + \frac{g}{c_{pd} T_0}\left(1 - e^{-\varepsilon z}\right) \left(\widehat{h_0} -  h_0^* \right).
\end{equation}
This equation is an analytic function of $B_{ud}$, $\varepsilon$, and the state variables within a sounding.  The first term on the RHS represents the direct dilution of the updraft's temperature perturbation via entrained air with no temperature perturbation, whereas the second term encapsulates the reduced condensation rate resulting from the entrainment of unsaturated air by the updraft, relative to an undiluted parcel. 

Before moving on to an analytic formula for ECAPE, we evaluate the accuracy of this analytic buoyancy formula by comparing the average buoyancy $\overline{B}$ between the level of free convection (LFC) and the level of neutral buoyancy (LNB) to that of the benchmark buoyancy profile and the formula from P20 (eqs. 4-5 therein\footnote{We also use the $B_{ud}$ computed with the benchmark parcel in the P20 formula to maximize this formula's accuracy.}).  Here, the LFC is the highest instance of zero buoyancy below the height of maximum buoyancy, and the LNB is the highest instance of zero buoyancy in the profile.  We define three metrics for evaluation: Pearson correlation coefficient $CC$ among soundings of $\overline{B}$ from eq. \ref{eq:banalytic} with $\overline{B}$ from the more accurate benchmark lapse rate formula; the fractional reduction in undiluted $\overline{B}$ by entrainment; and normalized root-mean-square-error (NRMSE), defined as the the average over all soundings of the squared difference between $\overline{B}$ from eq. \ref{eq:banalytic} and $\overline{B}$ from the benchmark lapse rate formula, divided by the magnitude of $\overline{B}$ from the benchmark formula.  These metrics are plotted as a function of $\varepsilon$ and updraft radius $R$ on the $x$ axis.  We relate $R$ to $\varepsilon$ using eq. \ref{eq:Avarepsilon}, with $k^2=0.18$, $P_r=\frac{1}{3}$, and $L_{mix}=120$ m following \citet{MPCS2021}. 
\begin{figure*}[!ht]
\centerline{\includegraphics[width=40pc]{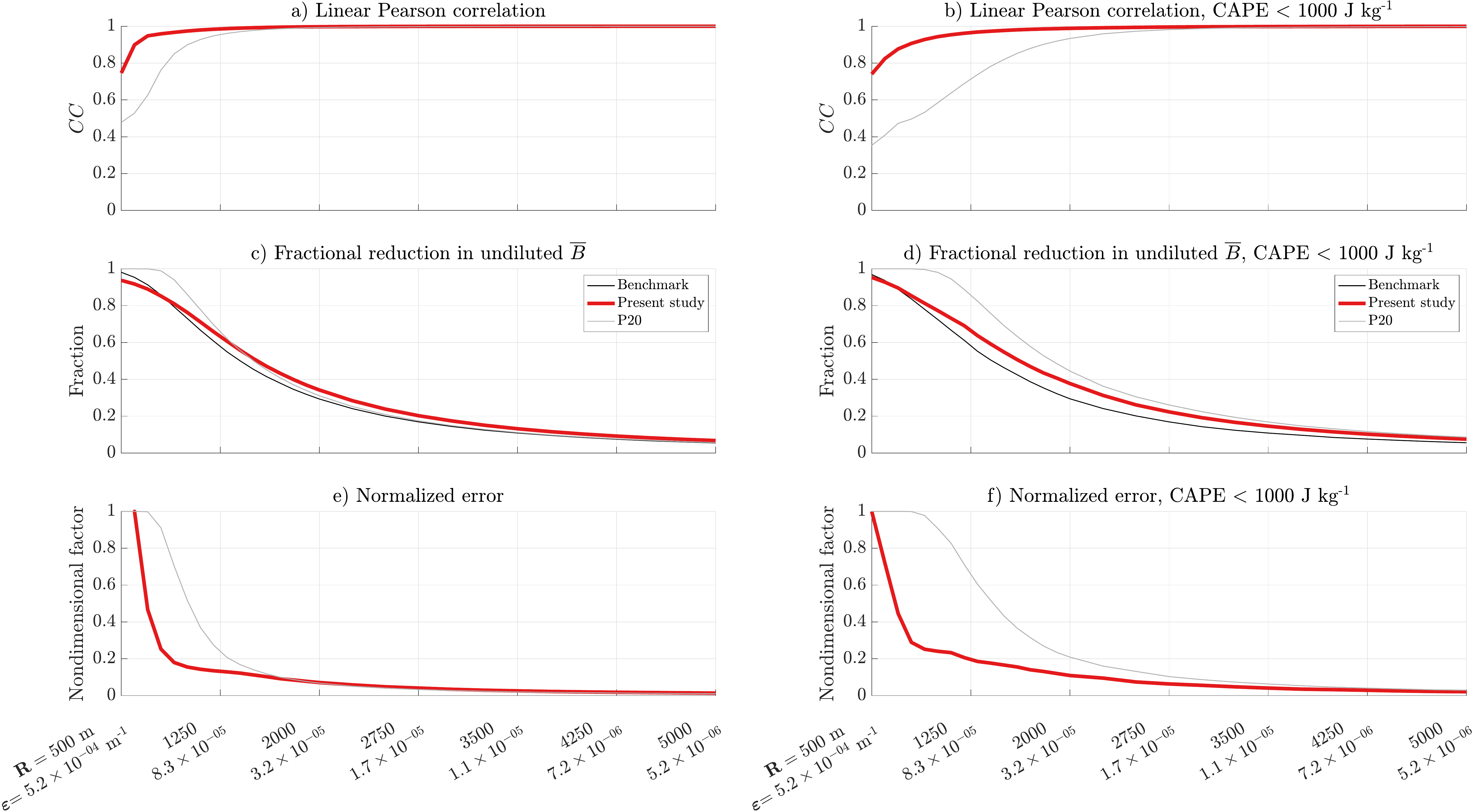}}
\caption{Comparison of vertically-averaged buoyancy $\overline{B}$ calculated using the formula from the present study (eq. \ref{eq:banalytic}, red), the P20 buoyancy formula (gray), and the benchmark parcel (black). Panels a,b show $CC$, c,d the fractional reduction in $\overline{B}$, and e,f the normalized error NRMSE. $CC$ and NRMSE are calculated relative to the benchmark parcel.
%Curves of $CC$ (panels a,b), fractional reduction in CAPE (panels c,d), and normalized error (panels e,f) for the vertically-averaged buoyancy $\overline{B}$ calculated using the formula from the present study (eq. \ref{eq:banalytic}, red), the P20 buoyancy formula (gray), and the benchmark parcel (black).  
Left panels show results from all \citet{THOM2003} soundings, and right panels show results from only soundings with $<$ 1000 J kg\textsuperscript{-1} undiluted CAPE.} \label{evalfig1}
\end{figure*}

The $CC$ of the new formula with the benchmark calculation is very close to 1 (Fig. \ref{evalfig1}a) for all $R>750$ m and for fractional reductions in CAPE of $<0.9$ (i.e., updrafts that realize 10 \% or more of their CAPE; Fig. \ref{evalfig1}c), which is the range of fractional reductions expected in midlatitude deep convection \citep[e.g.,][]{PNM2020,TEABT2021}.  For $R$ less than $750$ m and when fractional reductions approach 1, $CC$ begins to drop, suggesting that the formula is less accurate for strongly entraining weak convection.  The story is similar for NRMSE (Fig. \ref{evalfig1}e), which is relatively small in magnitude (i.e. $<0.1$) for $R>750$ m, but increases when $R$ falls below $750$ m.  Compared to the P20 formula, the new formula derived here has smaller NRMSE Fig. \ref{evalfig1}e) and larger $CC$ Fig. \ref{evalfig1}a), indicating that we have made an improvement in accuracy in the present derivation.  This improvement over the P20 formula is primarily due to an over-estimation of the fractional reduction in buoyancy via entrainment in the P20 formula that does not occur in the one derived here (Fig. \ref{evalfig1}c).  This difference is particularly noticeable when we restrict our analysis to soundings with less than $1000$ J kg$^{-1}$ of undiluted CAPE (Fig. \ref{evalfig1}b,d,f).  In this low CAPE regime, the NRMSE (Fig. \ref{evalfig1}f) and $CC$ (Fig. \ref{evalfig1}b) of the new formula are comparable to the errors for the whole sounding data set, whereas the P20 formula performs considerably worse with respect to both $CC$ and errors in the low CAPE regime.  

Our next task is to use eq. \ref{eq:banalytic} to obtain an expression for ECAPE. We formally define ECAPE as:
\begin{equation}
\label{eq:ECAPE}
\text{ECAPE} = \int_{z=LFC}^{z=LNB} B dz.
\end{equation}
Vertically integrating eq. \ref{eq:banalytic} from the LFC to the LNB and combining with eq. \ref{eq:ECAPE} yields:
\begin{equation}
\label{eq:ECAPE_analytic}
\text{ECAPE} = \int_{z=LFC}^{z=LNB}B_{ud} e^{-\varepsilon z}dz  + \int_{z=LFC}^{z=LNB}  \frac{g}{c_{pd} T_0}\left(1 - e^{-\varepsilon z}\right) \left(\widehat{h_0} -  h_0^* \right) dz.
\end{equation}
It will become advantageous later to have the integral bounds on the RHS of eq. \ref{eq:ECAPE_analytic} extend to the equilibrium level for an undiluted parcel\footnote{The equilibrium level is typically denoted with the acronym EL.  We instead use the symbol $H$ for compactness in equations.} $H$, rather than to the LNB.  We note that the integral of the first term from the LNB to the $H$ will always be positive, since $B_{ud}$ is positive below the $H$ by definition.  On the other hand, the integral of the second term over this range is typically negative (as will be discussed shortly), and at least partially cancels the contribution of the integral of the first term over this range.  Hence, we extend the upper bounds of these integrals to the $H$, assuming that the partial cancellation between the terms mitigates the resulting errors.

\begin{figure*}[!ht]
\centerline{\includegraphics[width=40pc]{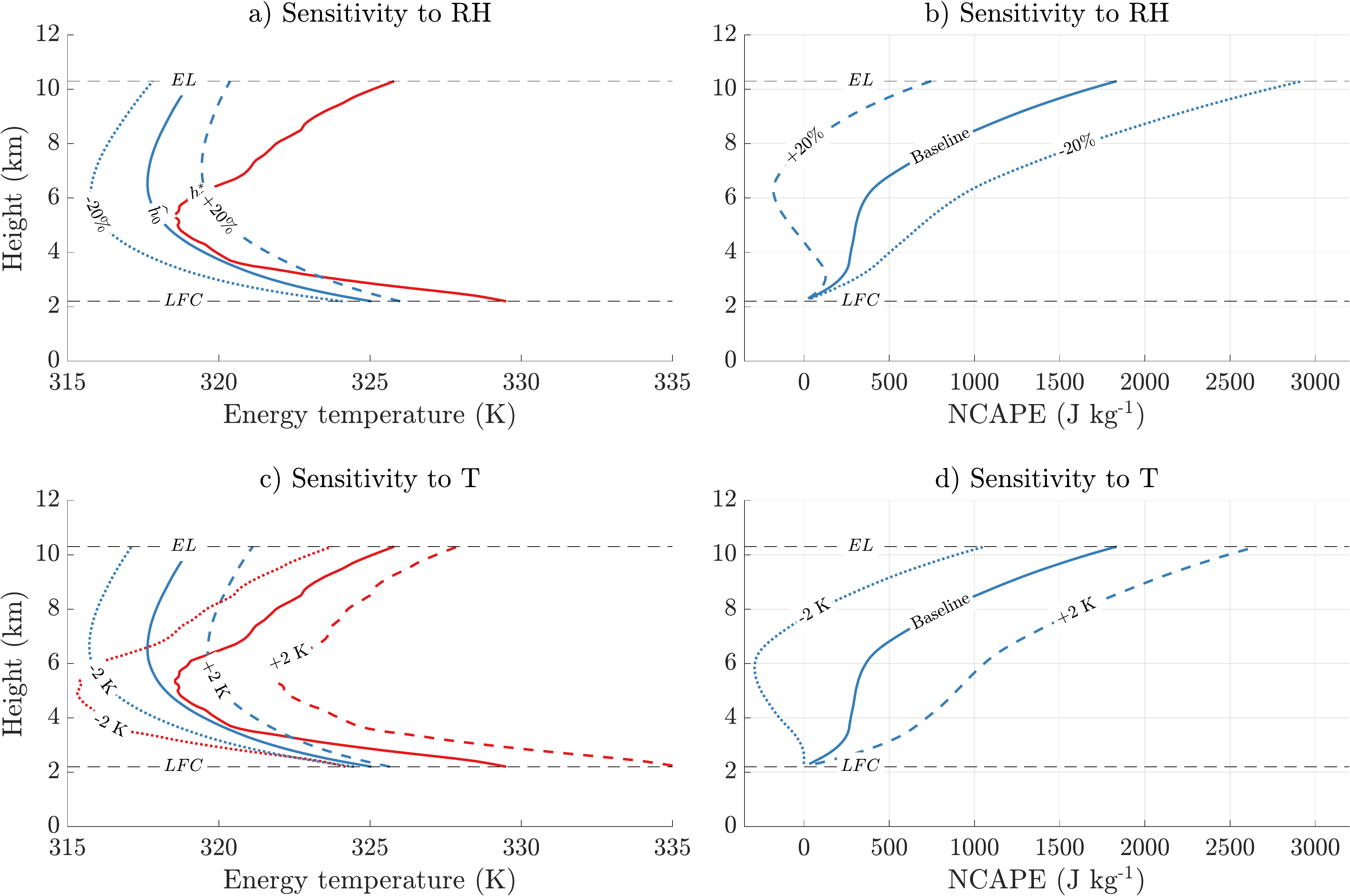}}
\caption{Demonstrations of the sensitivities of NCAPE to relative humidity (RH) and free tropospheric temperature.  Panel a: profiles of $h_0^*$ (red, divided by $c_{pd}$ to yield units of K), and $\widehat{h_0}$ (blue, K) for the baseline sounding (solid), RH increased by 20 \% (dashed blue), and RH decreased by 20 \% (dotted blue).  Panel b: profiles of NCAPE (J kg\textsuperscript{-1}) corresponding to panel a.  Panels c-d: analogous to panels a-b, but showing differences in $h_0^*$ and $\widehat{h_0}$ resulting from an increase in $T$ by 2 K with RH held constant (dashed), and a decrease in $T$ of 2 K with RH held constant (dotted).} \label{NCAPE}
\end{figure*}

To pull $\varepsilon$ out of the integrals in eq. \ref{eq:ECAPE_analytic}, we use integration by parts and these integral definitions to write the first term on the RHS of eq. \ref{eq:ECAPE_analytic} as:
\begin{equation}
\label{eq:Bexpan}
%old eq
% \int_{z=LFC}^{z=H}B_{ud} e^{-\varepsilon z}dz  = e^{-\varepsilon H} \text{CAPE} + \varepsilon \int_{z=LFC}^{z=H} e^{-\varepsilon z} \widehat{B}_z \left( z - LFC \right) dz
 \int_{z=LFC}^{z=H}B_{ud} e^{-\varepsilon z}dz  = e^{-\varepsilon H} \text{CAPE} + \varepsilon \int_{z=LFC}^{z=H} e^{-\varepsilon z} B_{ud} dz
\end{equation}
where
\begin{equation}
\label{eq:CAPE}
\text{CAPE} = \int_{z=LFC}^{z=H} B_{ud} dz.
\end{equation}
We make the approximation that $B_{ud}$ is linear with height on the RHS of eq. \ref{eq:Bexpan}:
\begin{equation}
\label{eq:bbar}
%\widehat{B}_z = \frac{1}{z - LFC}\int_{z^*=LFC}^{z^*=z} B_{ud} dz^*,
B_{ud} \approx \widehat{B_{ud}}\left(z - LFC\right),
\end{equation}
where $\widehat{B_{ud}}$ is the average undilute $B$ between the $LFC$ and $H$. We then vertically integrate eq. \ref{eq:Bexpan}, assume
% old text
%is the average of undilute $B$ between the LFC and a given height $z$.  
%To simplify the integral term in eq. \ref{eq:Bexpan}, we assume that $\widehat{B}_z$ is constant with height, assume 
that $LFC << H$ and hence $H - LFC \approx H$, and neglect entrainment below the LFC such that $e^{-\varepsilon LFC} \approx 1$.  We apply analogous assumptions to the 2nd term on the RHS of eq. \ref{eq:ECAPE_analytic}.  Modifying eq. \ref{eq:ECAPE_analytic} with these assumptions yields:
\begin{equation}
\label{eq:ECAPE_analytic2}
\text{ECAPE} = \left(\frac{1 - e^{-\varepsilon H}}{\varepsilon H}\right)\text{CAPE}  - \left(1 - \frac{1 - e^{-\varepsilon H}}{\varepsilon H}  \right) \text{NCAPE}
\end{equation}
where 
\begin{equation}
\label{eq:NCAPE}
\text{NCAPE} = - \int_{z=LFC}^{z=H} \frac{g}{c_{pd} T_0} \left(\widehat{h_0} - h_0^*\right) dz.
\end{equation}
NCAPE represents the buoyancy dilution potential of the free troposphere: the potential buoyancy loss that could be induced by entrainment mixing due principally to the saturation deficit of the environment. It is a purely environmental quantity that does not depend on parcel properties. As defined here with $\widehat{h_0}$, it specifically measures the energy difference between the saturation MSE at a given level and the mean MSE of the free troposphere below it. The latter captures the environment through which a parcel would have to rise, and potentially mix with, prior to reaching a particular level. Because $h_0^*$ is comparable to or larger than $\widehat{h_0}$ (Fig. \ref{NCAPE}a), NCAPE is typically (but not always) positive (Fig. \ref{NCAPE}b).  The difference term in the integral $\widehat{h_0} -  h_0^*$ (Fig. \ref{NCAPE}a) and hence the magnitude of NCAPE (Fig. \ref{NCAPE}b) will be larger when the free troposphere is dry and $\widehat{h_0}$ is far smaller than $h_0^*$, compared to when the free troposphere is moist and $\widehat{h_0}$ is closer in magnitude to $h_0^*$.  A warm free troposphere at a given RH generally increases the difference between $h_0^*$ and $\widehat{h_0}$ (Fig. \ref{NCAPE}c) compared to a situation when the free troposphere is cool at the same RH.  For a fixed RH, this makes NCAPE larger when the free troposphere is warm, relative to when it is cool (Fig. \ref{NCAPE}d).  Hence, NCAPE generally encapsulates the effects of tropospheric dryness and temperature on buoyancy via entrainment.

Eq. \ref{eq:ECAPE_analytic2} achieves the stated purpose of this derivation, since $\varepsilon$ is now outside of the integral terms. It will become advantageous in the next sub-section to further simplify the exponential terms in eq. \ref{eq:ECAPE_analytic2}.  One may consider making first order Taylor series approximations for the exponential terms.  For instance $ \frac{1 - e^{-\varepsilon H}}{\varepsilon H} \approx 1 - \varepsilon H$.  However, the exponential functions in eq. \ref{eq:ECAPE_analytic2} are strongly nonlinear with respect to $\varepsilon H$ in the range of $0 < \varepsilon H < 10$, which is the typical range we would encounter in our analysis, making the first order Taylor series approximation inaccurate (compare the blue and black lines in Fig. \ref{evalfig2}a).  Instead, we invert the exponential term $\frac{1 - e^{-\varepsilon H}}{\varepsilon H}$, approximate its inverse with a first order Taylor series, and then invert the result.  For instance:
\begin{figure*}[!ht]
\centerline{\includegraphics[width=40pc]{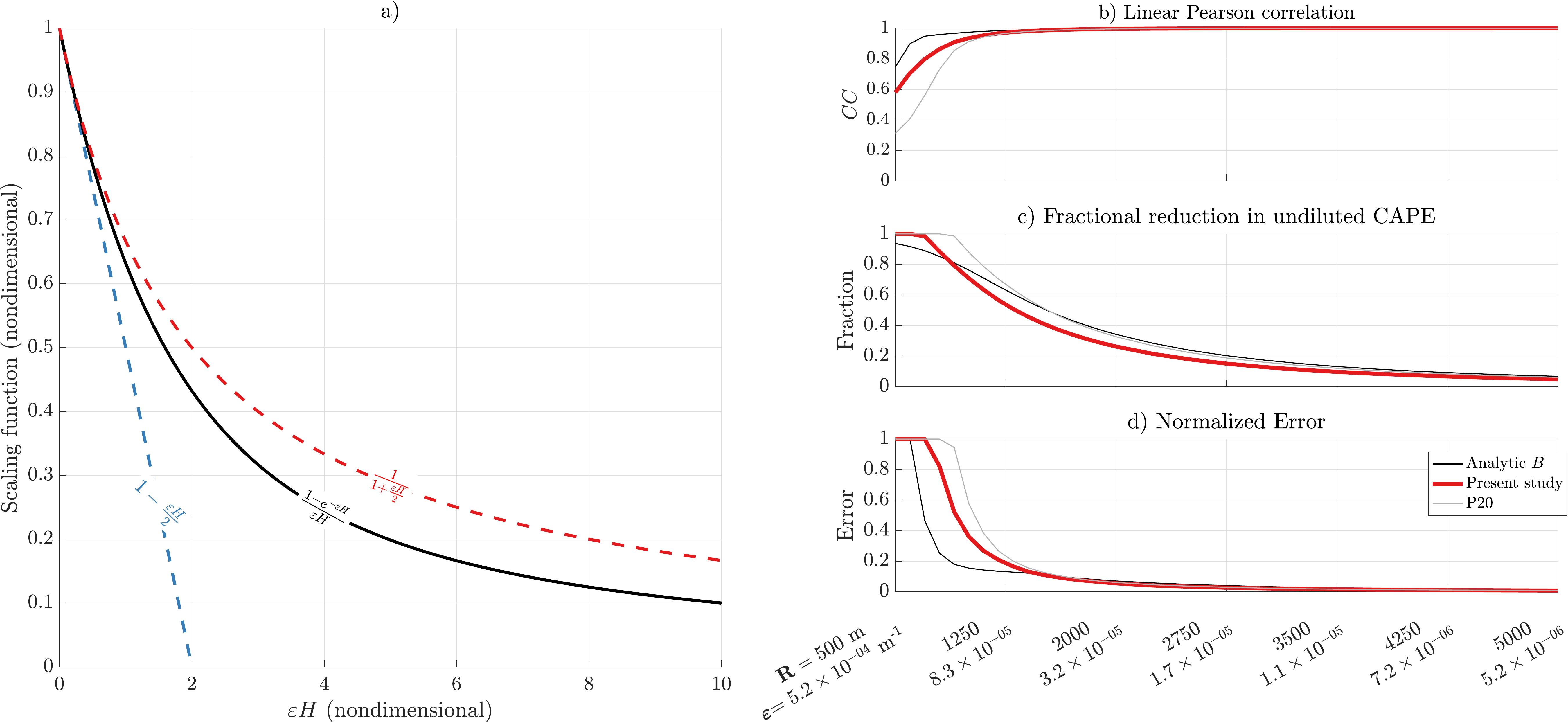}}
\caption{Panel a: comparison of the scale factor in eq. \ref{eq:ECAPE_analytic2} (solid black) with its first order Taylor series approximation (blue dashed), and the first order Taylor series approximation of its inverse (dashed red).  Panels b-d: analogous to Fig. \ref{evalfig1}a,b,c, but evaluating ECAPE from eq. \ref{eq:ECAPE_analytic3} (red, the present article), ECAPE from P20 (gray), and ECAPE from numerically integrating eq. \ref{eq:banalytic} (black), all relative to the benchmark calculation.} \label{evalfig2}
\end{figure*}
\begin{equation}
\label{eq:taylor_series}
\frac{\varepsilon H}{1 - e^{\varepsilon H}} \approx 1 + \frac{\varepsilon H}{2}.
\end{equation}
and consequently:
\begin{equation}
\label{eq:taylor_series2}
\frac{1 - e^{\varepsilon H}}{\varepsilon H} \approx \frac{1}{1 + \frac{\varepsilon H}{2}}.
\end{equation}
This approximation is far more accurate (compare the red and black lines in Fig. \ref{evalfig2}a).  Substituting these approximations into eq. \ref{eq:ECAPE_analytic2} and re-arranging yields:
\begin{equation}
\label{eq:ECAPE_analytic3}
\text{ECAPE} = \frac{\text{CAPE} - \frac{\varepsilon H}{2} \text{NCAPE}}{1 + \frac{\varepsilon H}{2}}.
\end{equation}
As a sanity check, examine the behavior of eq. \ref{eq:ECAPE_analytic3} under limiting scenarios.  For instance, in the limit of no entrainment where $\varepsilon \rightarrow 0$, $\text{ECAPE} \rightarrow \text{CAPE}$, which makes sense given that ECAPE for an undiluted parcel intuitively converges to the CAPE.  In the converse limit of $\varepsilon \rightarrow \infty$, we may use L'H\^{o}pital's rule to deduce that $\text{ECAPE} \rightarrow \text{NCAPE}$, which is inconsistent with the definition of CAPE as a quantity greater than or equal to zero.  However, this situation is easily remedied by simply setting ECAPE to 0 if eq. \ref{eq:ECAPE_analytic3}.  Finally, the case $\text{NCAPE}=0$ yields $\text{ECAPE} = \frac{CAPE}{1 + \frac{\varepsilon H}{2}}$, indicating that ECAPE is still smaller than CAPE when $\varepsilon \ne 0$ and hence dilution still reduces buoyancy in this situation. Indeed, for a saturated parcel to be positively buoyant in the first place requires $h > h_0^*$ (Eq. \ref{eq:bdefsat}), and since $h_0^* \ge h_0$ by definition, then $h>h_0$ and entrainment will dilute $h$ (e.g. Eq. \ref{eq:mseten}; and by extension, $B$). One specific example of this situation is an adiabatic atmosphere (dry or saturated; constant $h_0$), in which a parcel must be warmed in order to become positively buoyant and have non-zero CAPE, but in doing so the parcel will also have higher energy than the environment at all levels through which it rises.

The analytic formula for ECAPE in eq. \ref{eq:ECAPE_analytic3} loses a bit of accuracy relative to the numerically integrated analytic buoyancy equation at larger values of $\varepsilon$ (i.e., smaller updraft radii; Fig. \ref{evalfig2}b-d), but remains more accurate than the formula for maximum updraft vertical velocity $w_{max}$ from P20 (Eq. 18 therein), which is converted to ECAPE via $\frac{w_{max}^2}{2}$.  These errors stem from a slight underestimation of the fractional reduction in undiluted CAPE at large $\varepsilon$ values (Fig. \ref{evalfig2}c) that results from our changing of the integral bounds in eq. \ref{eq:ECAPE_analytic} from the LNB to $H$.  Despite these errors, this formula is quite accurate over the range of $R$ and $\varepsilon$ that typify deep moist convection (i.e., fractional reductions of no greater than 0.8, Fig. \ref{evalfig2}c).

\subsection{Relating fractional entrainment to environmental variables}

It will be convenient later in the derivation to manipulate a nondimensional form of eq. \ref{eq:ECAPE_analytic3}.  We define the nondimensional ECAPE as $\widetilde{E} \equiv \frac{\text{ECAPE}}{\text{CAPE}}$, the nondimensional NCAPE as $\widetilde{N} \equiv \frac{\text{NCAPE}}{\text{CAPE}}$, and the nondimensional fractional entrainment rate $\widetilde{\varepsilon} \equiv \varepsilon H$.  Using these definitions, we re-write eq. \ref{eq:ECAPE_analytic3} as:
\begin{equation}
\label{eq:ECAPE_NONDIM}
\widetilde{E} = \frac{1 - \frac{\widetilde{\varepsilon}}{2} \widetilde{N}}{1 + \frac{\widetilde{\varepsilon}}{2}}.
\end{equation}
Our next task is to eliminate $\widetilde{\varepsilon}$ from eq. \ref{eq:ECAPE_analytic3} by expressing this term as function of other updraft and environmental attributes.  We proceed by defining $\widetilde{R} \equiv \frac{R}{H}$ and use eq. \ref{eq:Avarepsilon} to write:
\begin{equation}
\label{eq:varepsilonnodim}
\widetilde{\varepsilon} = \epsilon \widetilde{R}^{-2},
\end{equation}
where
\begin{equation}
\label{eq:epsilon_const}
\epsilon = \frac{2k^2 L_{mix}}{ H P_r}.
\end{equation}
Combining eq. \ref{eq:varepsilonnodim} with eq. \ref{eq:ECAPE_NONDIM} yields:
\begin{equation}
\label{eq:ECAPE_NONDIM2}
\widetilde{E} = \frac{1 - \frac{\epsilon}{2\widetilde{R}^2} \widetilde{N}}{1 + \frac{\epsilon}{2\widetilde{R}^2}}.
\end{equation}

\begin{figure*}[!ht]
\centerline{\includegraphics[width=40pc]{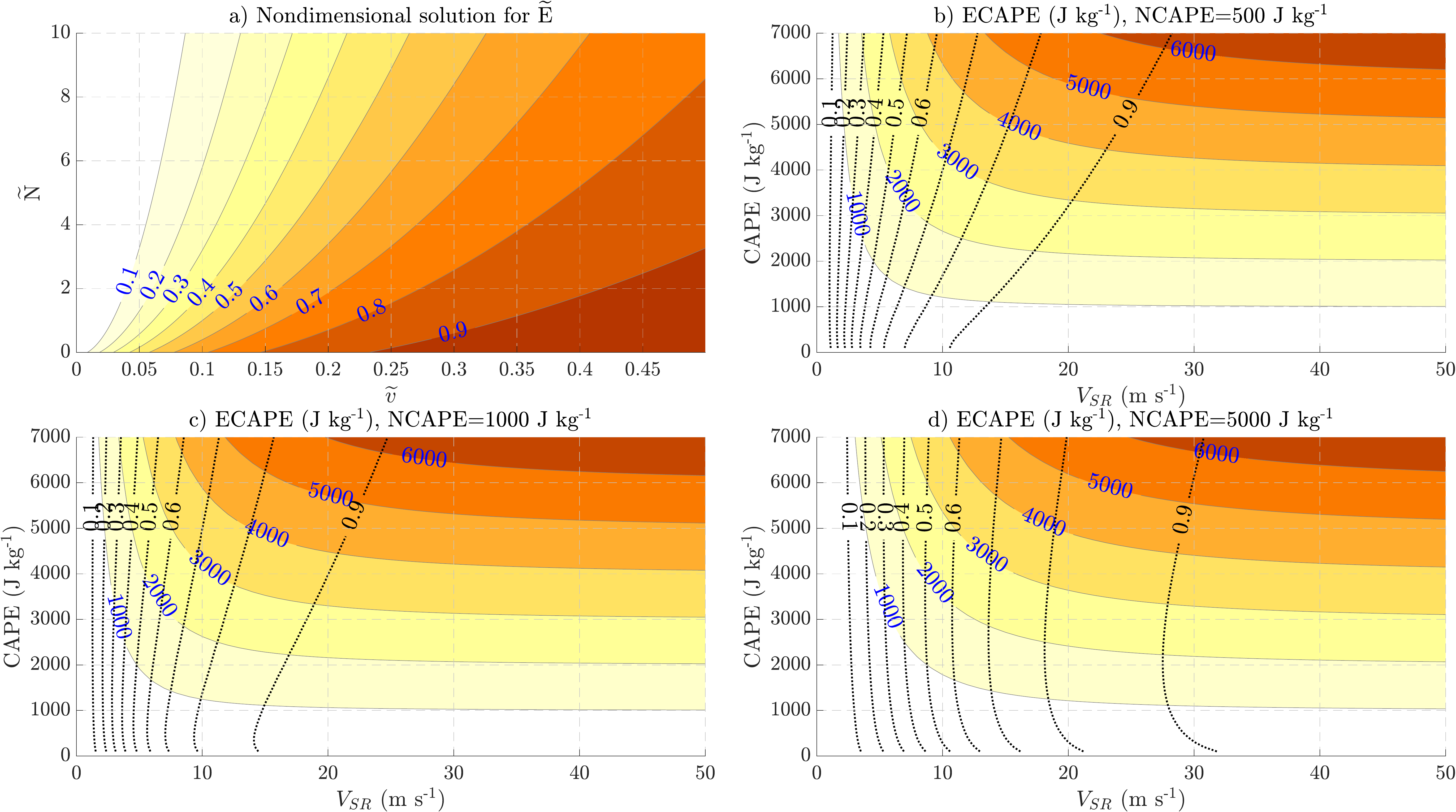}}
\caption{Panel a: $\widetilde{E}$ (shading) as a function of $\widetilde{v}$ ($x$ axis) and $\widetilde{N}$ ($y$ axis), with $H$ set to 12,000 m, $L=120$ m, $\alpha=0.8$, $\sigma = 1.131$, $k^2 = 0.18$, and $P_r = \frac{1}{3}$.  Panels b-d: ECAPE (shading, J kg\textsuperscript{-1}) as a function of $V_{SR}$ ($x$ axis, m s\textsuperscript{-1}) and undiluted CAPE ($y$ axis, J kg\textsuperscript{-1}), and $\widetilde{E}$ (black contours), with NCAPE = 500 J kg\textsuperscript{-1} (panel a), NCAPE = 1000 J kg\textsuperscript{-1} (panel b), and NCAPE = 5000 J kg\textsuperscript{-1} (panel c).  In panels b-d, $H$ is determined via $H = 5808 + 96.12 \sqrt{2CAPE}$, based on a linear regression between these variables among the soundings.  All other parameters are the same as in panel a.} \label{fs1}
\end{figure*}

Following P20 and \citet{PMNMMN21_1}, we may express $\widetilde{R}$ as a function of updraft and environmental attributes by making the following assumptions about updraft geometry and inflow:
\begin{enumerate}
    \item 
    Updrafts are cylindrical.
    \item
    Updraft radius $R$ is constant with height.  Numerous previous studies show this to be approximately valid \citep[e.g.,][]{SHER2013,DS2016,MPS2020}.
    \item
    We assume that all environmental storm-relative wind $\mathbf{V}_{SR}$ that encounters the cross-sectional area of the updraft on the upstream side becomes inflow.  Past studies also show this assumption to be reasonable \citep[e.g.,][]{PNM2019,PMNMMN21_2}.  
    \item
    The updraft maximum vertical velocity $w_{max}$ is proportional to the horizontally averaged vertical velocity $<w>$ at the same height, such that $<w> = \alpha w_{max}$, where $0 < \alpha < 1$ \citep[e.g.,][]{MOR2016c,MP2017}.
    \item
    The updraft maximum vertical velocity is primarily determined by updraft buoyancy, such that $w_{max} = \sqrt{2\text{ECAPE}}$.  This assumption is supported by \citep[][]{MP2017,J2017,PNM2019,PMNMT2020}.
    \item
    The maximum vertical velocity occurs at height $H$. 
\end{enumerate}
With these assumptions at hand, we start by writing the anelastic continuity equation in cylindrical coordinates as:
\begin{equation}
\label{eq:continuityn}
\rho_0 \frac{\partial r u  }{\partial r} + \rho_0 \frac{\partial v}{\partial \phi} + r \frac{\partial \rho_0 w }{\partial z}  = 0.
\end{equation}
Azimuthally integrating from $\phi=0$ to $\phi=2\pi$, radially integrating from $r=0$ to the updraft radius at $r=R$, and vertically integrating from the surface to $H$ (assuming $w= 0$ at $z=0$) and dividing by $2\pi$ yields:
\begin{equation}
\label{eq:continuity2}
H \widehat{\rho_0} \widehat{u}_R + R \frac{\rho_{0,H} < w_H >}{2} = 0.
\end{equation}
where 
\begin{equation}
\label{eq:uhatdef}
\widehat{u}_R = \frac{1}{2 \pi} \frac{\int_{z=0}^{z=H}  \rho_0 \int_{\phi=0}^{\phi=2 \pi} u d\phi dz}{\int_{z=0}^{z=H} \rho_0 dz}
\end{equation}
is the density-weighted vertical average of $u$ at radius $R$, and between the surface and height $H$, and represents the average inflow speed, 
\begin{equation}
\label{eq:wbardef}
< w > = \frac{1}{ \pi R^2} \int_{r=0}^{r=R} \int_{\phi=0}^{\phi=2 \pi} r w d\phi dr 
\end{equation}
is the area average of $w$ within radius $R$, $\widehat{\rho_0}$ is the vertical average of $\rho_0$ between the surface and height $H$, and $\rho_{0,H}$ is $\rho_0$ valid at height $H$.  Making use of $< w > = \alpha w_{max}$ (assumption 4) at height H and $\frac{w_{max}^2}{2} = ECAPE$ (assumption 5), and re-arranging eq. \ref{eq:continuity2} yields:
\begin{equation}
\label{eq:continuity3}
\widetilde{R} = - 2 \frac{\sigma}{\alpha} \frac{\widehat{u}_R}{\sqrt{2\text{ECAPE}}},
\end{equation}
where $\sigma = \frac{\widehat{\rho_0}}{\rho_{0,H}} > 1$. We may relate $\widehat{u}_R$ to the horizontal storm-relative wind speed $V_{SR} = | \mathbf{V}_{SR} |$, where $\mathbf{V}_{SR}$ is the storm-relative wind vector, by first defining the upstream flank of the updraft as the range from $\phi=- \frac{\pi}{2}$ to $\phi=\frac{\pi}{2}$. We next assume that all inflow is accomplished by the cloud-relative wind entering the upstream updraft flank, and the radial component of the environmental cloud-relative wind at the updraft edge is $u = - V_{SR} \cos{\phi}$ and $u=0$ m s\textsuperscript{-1} on the downstream edge.  These assumptions allow us to re-write eq. \ref{eq:uhatdef} as:
\begin{equation}
\label{eq:anglein}
\widehat{u}_R =  -\frac{1}{2 \pi} \frac{\int_{z=0}^{z=H}  \int_{\phi=-\frac{\pi}{2}}^{\phi=\frac{\pi}{2}} \rho_0 V_{SR} \cos{\phi} d\phi dz}{\int_{z=0}^{z=H} \rho_0 dz} = \frac{\widehat{V_{SR}}}{\pi},
\end{equation}
where $\widehat{V_{SR}}$ is the density weighted vertical average of $V_{SR}$ below height $H$.  In defining $\widetilde{v} \equiv \frac{\widehat{V_{SR}}}{\sqrt{2\text{CAPE}}}$, combining eqs. \ref{eq:continuity3} and \ref{eq:anglein} and the definition of $\epsilon$, and squaring and inverting the result, we obtain
\begin{equation}
\label{eq:continuity33}
\widetilde{R}^{-2} =  \frac{\alpha^2 \pi^2}{4 \sigma^2} \frac{ \widetilde{E}}{\widetilde{v}^2}.
\end{equation}
combining eq. \ref{eq:continuity33} with eq. \ref{eq:ECAPE_NONDIM2} to eliminate $R$ yields:
\begin{equation}
\label{eq:ECAPE_NONDIMBQ}
\widetilde{E}^2 \frac{\psi}{\widetilde{v}^2} + \widetilde{E}\left(1 + \frac{\psi}{\widetilde{v}^2} \widetilde{N} \right) - 1 = 0,
\end{equation}
where
\begin{equation}
\label{eq:psi}
\psi = \frac{ k^2 \alpha^2 \pi^2 L_{mix}}{4 P_r \sigma^2 H}.
\end{equation}
Solving for $\widetilde{E}$ using the quadratic formula gives:
\begin{equation}
\label{eq:ECAPE_NONDIM34}
\widetilde{E} = \frac{-1 - \frac{\psi}{\widetilde{v}^2} \widetilde{N} + \sqrt{\left(1 + \frac{\psi}{\widetilde{v}^2} \widetilde{N} \right)^2 + 4\frac{\psi}{\widetilde{v}^2} }}{2 \frac{\psi}{\widetilde{v}^2}},
\end{equation}
where we have neglected the negative quadratic root that yields an imaginary solution.  Solutions for $\widetilde{E}$, which represent the fractional reduction of undiluted CAPE by entrainment, are contoured in Fig. \ref{fs1}a  as a function of $\widetilde{v}$ (non-dimensional storm-relative flow speed) and $\widetilde{N}$ (non-dimensional NCAPE).  In general, $\widetilde{E}$ increases from left-to-right in the figure as $\widetilde{v}$ becomes large, indicating stronger storm-relative inflow, wider updrafts, and hence smaller fractional entrainment.  From bottom-to-top on the figure, $\widetilde{E}$ decreases as $\widetilde{N}$ increases.  This trend occurs because larger $\widetilde{N}$ implies a drier and/or warmer mean free troposphere, both of which amplify entrainment-driven dilution relative to situations with a cooler and/or moister free troposphere.

In dimensional form, eq \ref{eq:ECAPE_NONDIM34} is:
\begin{equation}
\label{eq:ECAPE_DIM34}
\text{ECAPE} =  \frac{-1 - \frac{2\psi}{V_{SR}^2} \text{NCAPE} + \sqrt{\left(1 + \frac{2\psi}{V_{SR}^2} \text{NCAPE}\right)^2 + \frac{8\psi}{V_{SR}^2} \text{CAPE}}}{4 \frac{\psi}{V_{SR}^2}}.
\end{equation}
Solutions for ECAPE from eq. \ref{eq:ECAPE_DIM34} as a function of $V_{SR}$ and CAPE are shown in Fig. \ref{fs1}b,c,d for NCAPE=500 J kg\textsuperscript{-1}, 1000 J kg\textsuperscript{-1}, and 5000 J kg\textsuperscript{-1} respectively.  In general, curves of ECAPE take on hyperbolic shapes with respect to the $x$ and $y$ axes, with contours of ECAPE parallelling the $x$ axis for large $V_{SR}$, and the $y$ axis for small $V_{SR}$ and large CAPE, and with the largest values coinciding with the largest $V_{SR}$ and undiluted CAPE in the upper-right corners of the figures.  This pattern means that different combinations of $V_{SR}$ and undiluted CAPE may result in similar ECAPE.  For instance, an environment with 1000 J kg\textsuperscript{-1} of undiluted CAPE, a $V_{SR}$ of 30 m s\textsuperscript{-1}, and an NCAPE of -5000 J kg\textsuperscript{-1}, has an ECAPE of roughly 1000 J kg\textsuperscript{-1} (Fig. \ref{fs1}d).  Mature isolated deep convective updrafts in this environment will be sufficiently wide, due to their large $V_{SR}$, such that their cores are approximately undiluted and they realize nearly all of their undiluted CAPE.  A contrasting environment with 6000 J kg\textsuperscript{-1} of undiluted CAPE and an NCAPE of -5000 J kg\textsuperscript{-1}, but with a $V_{SR}$ of only 5 m s\textsuperscript{-1} will have a similar ECAPE of 1000 J kg\textsuperscript{-1}.  Despite the large undiluted CAPE in the second environment, updrafts are narrow and substantially diluted by entertainment because of small $V_{SR}$.

Consistent with the dependence of $\widetilde{E}$ on $\widetilde{N}$ seen in Fig. \ref{fs1}a, the fractional reduction in undiluted CAPE by ECAPE increases as NCAPE increases, particularly for smaller values of undiluted CAPE.  This is most evident as a movement to the right of the contours of $\widetilde{E}$ (black) in Fig. \ref{fs1}b-d as NCAPE increases, indicating that an updraft with a given combination of undiluted CAPE and $V_{SR}$ will realize less of its CAPE when NCAPE is large, compared to when NCAPE is small.  

\subsection{Accounting for kinetic energy the storm derives from its environment}

While it is somewhat infrequent, past studies have documented instances in supercells where the maximum updraft $w$ exceeds $\sqrt{2\text{CAPE}}$ for extended periods of time \citep[e.g.,][]{Fiedler1994}.  Hence, there are factors, such as vertical pressure gradient accelerations, that can explain why updrafts are sometimes more intense than buoyancy alone would suggest. This section introduces a simple adjustment factor to the ECAPE formula to represent of how such pressure effects redirect environmental kinetic energy into the updraft.  To derive this adjustment factor, we must make the following assumptions:
\begin{enumerate}
    \item 
    The Lagrangian evolution of kinetic energy following an air parcel is well described by the Boussinesq approximation, meaning that $\rho_0$ is constant.  Past studies have shown that errors related to an over-estimation of $\rho_0$ aloft in deep convective environments have a small effect on analytic solutions for vertical velocity, \citep[e.g.,][]{MOR2015a,MOR2015b}.
    \item
    Perturbation pressure accelerations in the middle-to-upper troposphere are neglected.  Pressure perturbations aloft may be large, but they typically occur within the toroidal circulations of moist thermals \citep[e.g.,][]{RC2015,MP2017,PC2021}.  As parcels ascend through these thermals, they experience an upward acceleration below the minimum in $p'$, and then a commensurate downward acceleration above the minimum in $p'$.  Hence, any temporary $KE$ gained by the interaction of a parcel with these pressure perturbations is quickly lost.  We therefore neglect pressure perturbations at the height of maximum $w$.
    \item
    Direct dilution of $KE$ via entrainment is negligible.  This assumption is also supported by past studies \citep[e.g.,][]{SHER2013}.  Note that entrainment will still indirectly affect KE via the entrainment-driven dilution of updraft buoyancy.
    \item
    Updrafts are approximately steady, such that $\frac{\partial}{\partial t}$ of quantities are small.
    \item
    The magnitude of convective inhibition (CIN) is negligable relative to the magnitude of ECAPE.
    \item
    Horizontal storm-relative flow vanishes at the height of $w_{max}$.
\end{enumerate}

We may use the first assumption to write eq. 15 in \citet{PC2021}, which describes the Lagrangian tendency for $KE$, as as:
\begin{equation}
\label{eq:KE}
\frac{d KE}{dt} =  \mathbf{V} \cdot \nabla \left( \frac{p'}{\rho_0} \right) + wB
\end{equation}
where $p'$ is a pressure perturbation.  We define $KE$ here in an updraft relative sense, such that $KE = \frac{u_{CR}^2 + v_{CR}^2 + w^2}{2}$, where $u_{CR}$ and $v_{CR}$ are the $u$ and $v$ cloud-relative wind components.  Because of the steady state assumption, we may substitute $\frac{d}{dt} \left(\frac{p'}{\rho_0} \right)= \mathbf{V} \cdot \nabla \left( \frac{p'}{\rho_0} \right) $.  We further use the chain rule to write $\frac{d}{dt} = w \frac{d}{dz}$, where $\frac{d}{dz}$ is the rate of change of a quantity as a parcel changes height.  Making these assumptions and substitutions, and integrating from a parcel starting position (defined as $z=0$) to an ending position at the height of $w_{max}$ yields the following form of the classical Bernoulli equation:
\begin{equation}
\label{eq:KE2}
KE_{LNB} - KE_{0} = \frac{p'_{LNB}}{\rho} - \frac{p'_{0}}{\rho} + \int_{z=0}^{z=LNB} B dz.
\end{equation}
If a parcel originates within an updraft's unmodified background environmental flow then $p'=0$, $w=0$, and $KE_0 = \frac{V_{SR}^2}{2} $.  We may also neglect $\frac{p'_{LNB}}{\rho}$ because of assumption (2) above. Finally, we note that $\int_{z=0}^{z=LNB} B dz = $ ECAPE + ECIN, where ECIN is the convective inhibition for an entraining parcel (ECAPE here is defined via eq. \ref{eq:ECAPE_DIM34}).  Combining all these assumptions and substitutions, neglecting ECIN, and assuming that horizontal storm-relative flow vanishes at the height of $w_{max}$ gives:
\begin{equation}
\label{eq:KE4}
\text{ECAPE}_A = \frac{w_{max}^2}{2} = \frac{V_{SR}^2}{2} + \text{ECAPE}
\end{equation}
where the subscript $A$ indicates ``adjusted''.  According to this equation, the role of low-level pressure perturbations is to preserve the incoming cloud-relative horizontal kinetic energy, deflecting it into the vertical.  Further, the maximum updraft kinetic energy at the height of $w_{max}$ consists of the sum of the kinetic energy gained from the release of ECAPE and the kinetic energy of the redirected inflow.  Nondimensionalizing by the undiluted CAPE yields:
\begin{equation}
\label{eq:KE4}
\widetilde{E}_A = \widetilde{v}^2 + \widetilde{E},
\end{equation}
where $\widetilde{E}_A$ is the nondimensional analogy to ECAPE$_A$.  Recall that in the derivation in the previous sub-section, we neglected pressure effects and assumed that $\text{ECAPE} = \frac{w_{max}^2}{2}$ when deriving the expression for $R^{-2}$ in eq. \ref{eq:continuity33}.  Now we must account for the influence of the added contribution to $w_{max}$ from velocity from environmental kinetic energy on updraft radius.  Hence, we set $\text{ECAPE}_A = \frac{w_{max}^2}{2}$, and adjust eq. \ref{eq:continuity33} using eq. \ref{eq:KE4} to:
\begin{equation}
\label{eq:continuity333}
\widetilde{R}^{-2} =  \frac{\alpha^2 \pi^2}{4\sigma^2} \frac{w_{max}^2}{V_{SR}^2} =  \frac{\alpha^2 \pi^2}{4 \sigma^2} \left( \frac{\widetilde{E}}{\widetilde{v}^2} + 1 \right).
\end{equation}
Combining eqs. \ref{eq:KE4}-\ref{eq:continuity333} with eq. \ref{eq:ECAPE_NONDIM2} yields:
\begin{equation}
\label{eq:ECAPEA_NONDIMBQ}
\widetilde{E}^2 \frac{\psi}{\widetilde{v}^2} + \widetilde{E}\left(1 + \psi + \frac{\psi}{\widetilde{v}^2} \widetilde{N}\right) - 1 + \psi \widetilde{N}= 0,
\end{equation}
Solving $\widetilde{E}$ using the quadratic formula and then plugging the result into eq. \ref{eq:KE4} to solve for $\widetilde{E}_A$ gives:%%
\begin{equation}
\label{eq:ECAPE_NONDIM_DYNAM}
\widetilde{E}_A = \widetilde{v}^2 + \frac{-1 - \psi - \frac{\psi}{\widetilde{v}^2} \widetilde{N} + \sqrt{\left(1 + \psi + \frac{\psi}{\widetilde{v}^2}\widetilde{N} \right)^2 + 4\frac{\psi}{\widetilde{v}^2} \left( 1 - \psi\widetilde{N}\right) }}{2 \frac{\psi}{\widetilde{v}^2}},
\end{equation}
which may be written dimensionally as:
\small
    \begin{equation}
    \label{eq:ECAPE_DIM_DYNAM}
    \begin{aligned}
    \text{ECAPE}_A = \frac{V_{SR}^2}{2} + \frac{-1 - \psi - \frac{2 \psi}{V_{SR}^2} \text{NCAPE}}{4 \frac{\psi}{V_{SR}^2}} + \\ \frac{\sqrt{\left(1 + \psi + \frac{2\psi}{V_{SR}^2} NCAPE \right)^2 + 8\frac{\psi}{V_{SR}^2} \left( \text{CAPE} - \psi \text{NCAPE}\right) }}{4 \frac{\psi}{V_{SR}^2}}.
    \end{aligned}
    \end{equation}
    \normalsize

\begin{figure*}[!ht]
\centerline{\includegraphics[width=40pc]{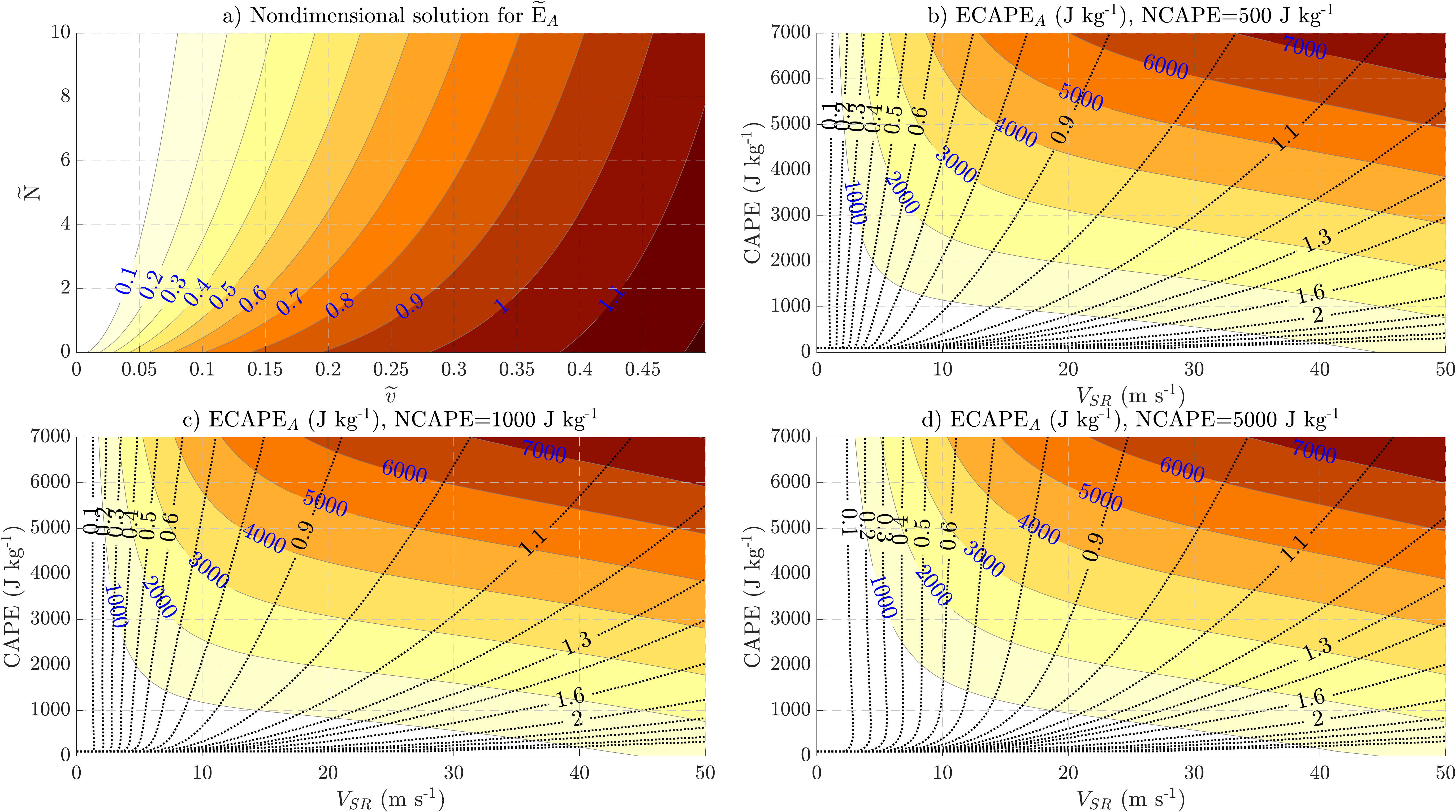}}
\caption{Same as Fig. \ref{fs2}, but showing $\widetilde{E}_A$ (panel a), and ECAPE$_A$ (panels b-d).} \label{fs2}
\end{figure*}

The solution for $\widetilde{E}_A$ from eq. \ref{eq:ECAPE_DIM_DYNAM} (Fig. \ref{fs2}a) is similar to that of $\widetilde{E}$ from eq. \ref{eq:ECAPE_NONDIM34} at small values of $\widetilde{v}$, but diverges notably from $\widetilde{E}$ at large $\widetilde{v}$, exceeding 1 (indicating that ECAPE\textsubscript{A} surpasses CAPE).  Similar behavior is evident in the solutions for ECAPE\textsubscript{A} as a function of $V_{SR}$ and CAPE (Fig. \ref{fs2}b-d).  Notably, ECAPE\textsubscript{A} is similar to ECAPE at smaller values of $V_{SR}$, but larger than ECAPE\textsubscript{A} at large values of $V_{SR}$, which is evident as a persistent downward slant of ECAPE\textsubscript{A} as one moves from left-to-right on the figure.  Again, we see that drastically different combinations of $V_{SR}$ and CAPE can yield the same value of ECAPE\textsubscript{A}.  For instance, an environment with NCAPE of 500 J kg\textsuperscript{-1},  1000 J kg\textsuperscript{-1} of CAPE, and a $V_{SR}$ of 45 m s\textsuperscript{-1} will have an ECAPE\textsubscript{A} of 2000 J kg\textsuperscript{-1}.  A starkly contrasting environment with NCAPE of 5000 J kg\textsuperscript{-1},  7000 J kg\textsuperscript{-1} of CAPE, and a $V_{SR}$ of 7 m s\textsuperscript{-1} will also have an ECAPE\textsubscript{A} of 2000 J kg\textsuperscript{-1}.

To illustrate the circumstances under which pressure accelerations (as they have been formulated here) have the greatest enhancement effect on updrafts, we examine the quantity $F = \sqrt{\frac{\text{ECAPE}_A}{\text{ECAPE}}} - 1	$, which is equal to the ratio of the fractional enhancement in $w_{max}$ due to pressure accelerations.  Fractional enhancement is quite small ($<0.1$) for most combinations of $V_{SR}$ and CAPE.  It only becomes larger than 0.1 for smaller values of CAPE and/or larger values of $V_{SR}$.  Physically, when CAPE is large and/or $V_{SR}$ is small, the kinetic energy generation from buoyancy dominates the updraft kinetic energy budget.  Whereas, when CAPE is small and/or $V_{SR}$ is large, the kinetic energy input from the environmental wind becomes comparable to the kinetic energy generation from buoyancy.  Given this distribution of $F$, a potential explanation for why many past studies have found that $w_{max}$ is primarily determined by buoyancy is that the CAPE and $V_{SR}$ in these simulations fell within the region of the parameter space where $F$ is small.  In other words, the kinetic energy input into the updraft via the background environmental flow is insignificant compared to the kinetic energy generation via the release of CAPE in most storm environments.

\begin{figure*}[!ht]
\centerline{\includegraphics[width=30pc]{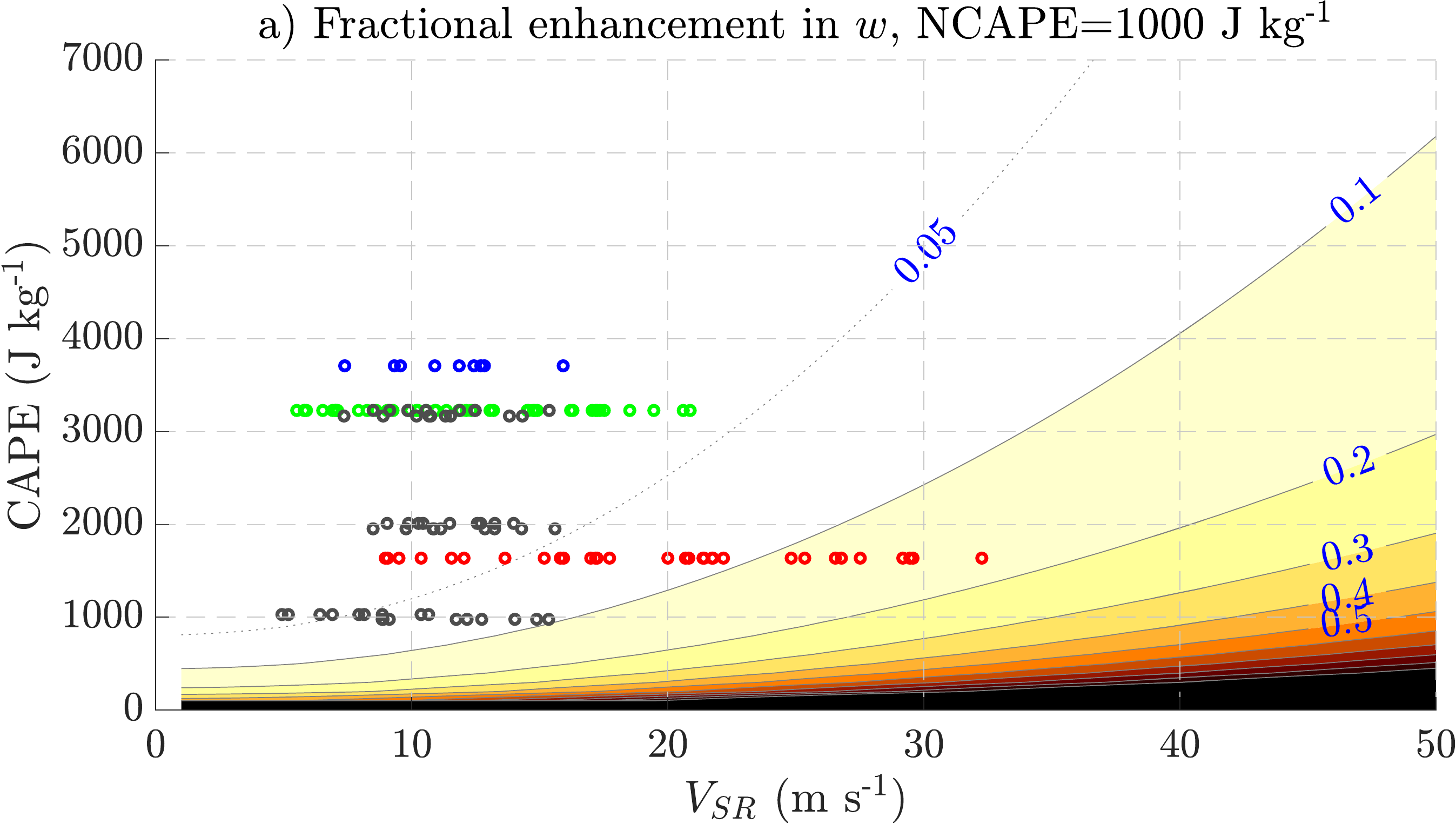}}
\caption{$F$ (shading, nondimensional) as a function of $V_{SR}$ ($x$ axis, m s\textsuperscript{-1}) and CAPE ($y$ axis, J kg\textsuperscript{-1}).  Colored dots indicate the $V_{SR}$ and CAPE from the simulated storms analyzed in section 4.} \label{pres}
\end{figure*}

\begin{figure*}[!ht]
\centerline{\includegraphics[width=40pc]{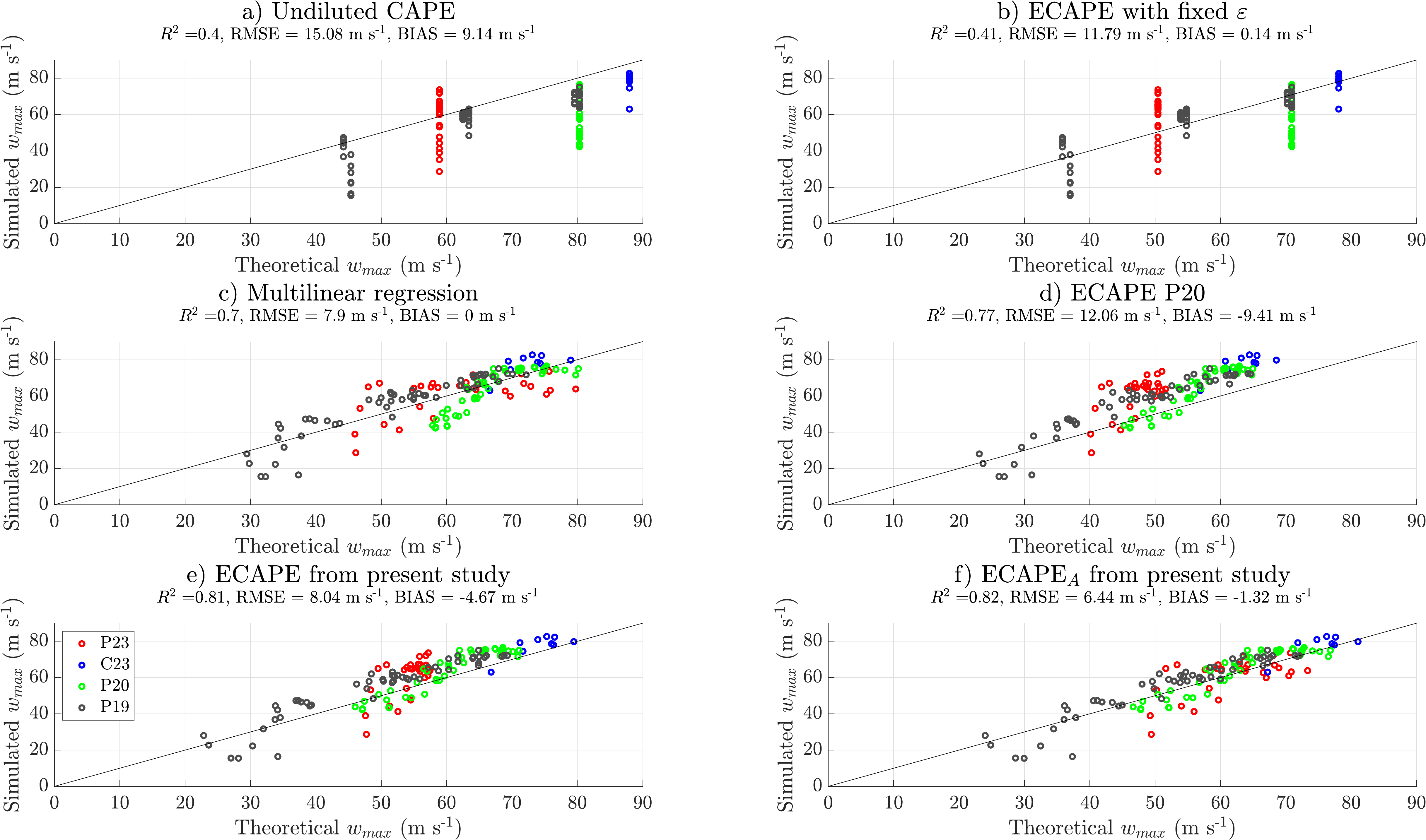}}
\caption{All panels: predicted $w_{max}$ ($x$ axis, m s\textsuperscript{-1}) versus simulated $w_{max}$ ($y$ axis, m s\textsuperscript{-1}).  Predictors are: the traditional ``thermodynamic speed limit" $\sqrt{2\text{CAPE}}$ (panel a), ECAPE with the fixed $\varepsilon$ that minimized the RMSE (panel b), a multi-linear regression with $V_{SR}$ and $\sqrt{2\text{CAPE}}$ as predictors (panel c), ECAPE from P20 (panel d), ECAPE from the present study (panel e), and ECAPE\textsubscript{A} from the present study (panel f).  Bias, RMSE and $R^2$ values are shown in the title of each plot.  Colors correspond to the study where the simulations originated (see the legend in panel e).} \label{verf}
\end{figure*}

\section{Evaluation of the formulas}

\subsection{Comparison of predicted $w_{max}$ with the output from past simulations}

We will compare the formula's predictions to the vertical velocities from simulations to evaluate the ECAPE and ECAPE$_A$ formulas.  The simulations, which featured a mix of supercells and multicellular clusters, originate from four past studies: \citet{CEA2023} (C23, 9 simulations), \citet{PEA2023} (P23, 32 simulations), \citet{PNM2020b} (P20, 48 simulations), and \citet{PNM2019} (54 simulations).  All simulations used Cloud Model 1 \citep[CM1][]{BF2002} and were initialized with soundings that featured a variety of different wind and thermodynamic profiles.  Horizontal grid spacing was 100 m in P23 and C23, and 250 m in P20, and P19.  Vertical grid spacing was 100 m or less in the troposphere in all simulations.  Additional details of the model configurations are omitted here to save room, but are available in the studies referenced in this paragraph.  

We computed all subsequent quantities with the initial model thermodynamic and wind profiles and storm motions in past simulations. Predictions of $w_{max}$ were derived by taking the square root of half of the predicted CAPE and ECAPE values.  We compared the predicted values of $w_{max}$ to the median $w_{max}$ during the 1-3 hour time range in the simulations, excluding tornadic periods in the P23 and C23 simulations (see those studies for definitions of ``tornadic periods").  The parameter $V_{SR}$ was computed by subtracting the tracked motion vector of simulated updrafts from the initial model profile, and averaging the resulting storm-relative wind profile in the 0-1 km layer.  Other layer averages, including 0-500 m, 0-2 km, 0-3 km, and the density weighted average from the surface to the EL gave nearly identical results. 

We will first see how well $\sqrt{2\text{CAPE}}$, which is the traditional ``thermodynamic speed limit'', predicts $w_{max}$ (Fig. \ref{verf}a).  This parameter loosely captures the differences in $w_{max}$ among groups of simulations, but does not capture any of the variability in $w_{max}$ among simulations that shared the same CAPE.  Most $w_{max}$ were less than the traditional thermodynamic speed limit (i.e., below the 1-to-1 line).  However, the bulk of the P23 simulations and a few of the P19 simulations exceeded this threshold, by up to 15 m s\textsuperscript{-1}.  The $V_{SR}$ and CAPE of these simulations puts them in the portion of the parameter space where our theoretical representation of pressure effects predicts that their $w_{max}$ should exceed $\sqrt{2\text{CAPE}}$ (see the gray and red dots in Fig. \ref{pres}).  The coefficient of determination ($R^2$) of $\sqrt{2\text{CAPE}}$ with simulated $w_{max}$ was 0.38, with a root-mean-square-error (RMSE) of roughly 15 m s\textsuperscript{-1}.

To see if we can do a better job of predicting $w_{max}$ with ECAPE that uses a fixed entrainment rate, we found the $\varepsilon$ that yielded the smallest RMSE between predictions by eq. \ref{eq:ECAPE_analytic2} and simulated $w_{max}$ (this value was $\varepsilon=2.25\times10^{-5}$ m\textsuperscript{-1}).  This prediction reduces the RMSE to 12.2 m s\textsuperscript{-1}, but does not improve the $R^2$ much (Fig. \ref{verf}b).  Hence, with no knowledge of how the variations in environmental wind profiles affect entrainment, ECAPE with a fixed entrainment rate only slightly improves predictions of the mean $w_{max}$ among groups of simulations, but does not capture any of the variance in $w_{max}$ within a particular group.

We \textit{can} do a better job of predicting $w_{max}$ by forming a mult-linear regression with $\sqrt{2\text{CAPE}}$ and $V_{SR}$ as predictors, and $w_{max}$ as a predictand.  This regression equation takes the form $w_{max,pred} = 0.7823 \sqrt{2\text{CAPE}} + 1.503 V_{SR} - 13.3437$.  The predictions by this formula reduce RMSE to 7.95 m s\textsuperscript{-1} and increase the $R^2$ to 0.7 (Fig. \ref{verf}c).  This formula also produces an improved subjective correspondence between predicted and simulated $w_{max}$.

The ECAPE formula from P20, computed using all the procedures and parameter values described in that study, also better captures the variability in $w_{max}$ among simulations with the same CAPE value than the $\sqrt{2\text{CAPE}}$ and ECAPE with a fixed entrainment rate, with a $R^2$ with $w_{max}$ of 0.71.  The RMSE of 13 m s\textsuperscript{-1}, however, is inferior to that of the linear regression and comparable to that of $\sqrt{2\text{CAPE}}$ and ECAPE with a fixed entrainment rate.  This large error stems from a low bias in predictions from this formula, relative to the values in simulations, which is demonstrated by the dots mostly falling to the left of the one-to-one line in Fig. \ref{verf}b).  Recall that P20 used a $\varepsilon \sim R^{-1}$ scaling, and the buoyancy formula from that study consequently over-estimated the fractional reduction in undiluted buoyancy by entrainment.  Both of these factors may have contributed to the formula's bias.

To evaluate the ECAPE and ECAPE\textsubscript{A} derived in the present study, we set $L_{mix}=120$ m when evaluating the ECAPE formulas derived in the present study against the P23 and C23 simulations, and $L_{mix}=250$ m when evaluating against the P20, N20, and P19 simulations to account for their coarser grid spacing.  All other parameter values were the same as those used to generate Figs. \ref{fs1}-\ref{fs2}.  The new ECAPE formula improves correspondence ($R^2=0.79$), reduces the low bias in prediction, and substantially decreases RMSE (8.2 m s\textsuperscript{-1}) relative to the formula from P20 and the linear regression.  Dots in Fig. \ref{verf}c fall close to the 1-1 line, suggesting that the $\varepsilon \sim R^{-2}$ scaling better reflects the trends in entrainment-driven dilution in the simulations than $\varepsilon \sim R^{-1}$.

The ECAPE\textsubscript{A} formula further improves correspondence between predicted and simulated $w_{max}$ ($R^2=0.82$), decreases RMSE to 6.4 m s\textsuperscript{-1}, and brings points closer to the 1-to-1 line.  The most notable difference between ECAPE\textsubscript{A} and ECAPE occurs with the P23 simulations, whose $w_{max}$ substantially exceeded $\sqrt{2 \text{CAPE}}$ (red dots above the 1-to-1 line in Fig. \ref{verf}a) and was under-predicted by the ECAPE formulas from both P20 (red dots above the 1-to-1 line in Fig. \ref{verf}b) and the present study (red dots above the 1-to-1 line in Fig. \ref{verf}c).  The ECAPE\textsubscript{A} brings the red dots much closer to the 1-to-1 line, correctly reflecting that $w_{max}$ in many of these simulations exceeded $\sqrt{2 \text{CAPE}}$. 

The take home message is that the two formulas derived in the present study are superior predictors of $w_{max}$ when compared to CAPE and ECAPE with a fixed entrainment rate.  They also perform better than a simple linear regression that includes CAPE and $V_{SR}$, suggesting that the additional information contained in our formula about the environmental thermodynamic profile via the NCAPE parameter is critical to accurately representing the effects of entrainment on $w_{max}$.  Finally, the new ECAPE formulas correct a low bias in the older P20 formula.

\subsection{Properties of ECAPE in severe weather proximity soundings}

\begin{figure*}[!ht]
\centerline{\includegraphics[width=40pc]{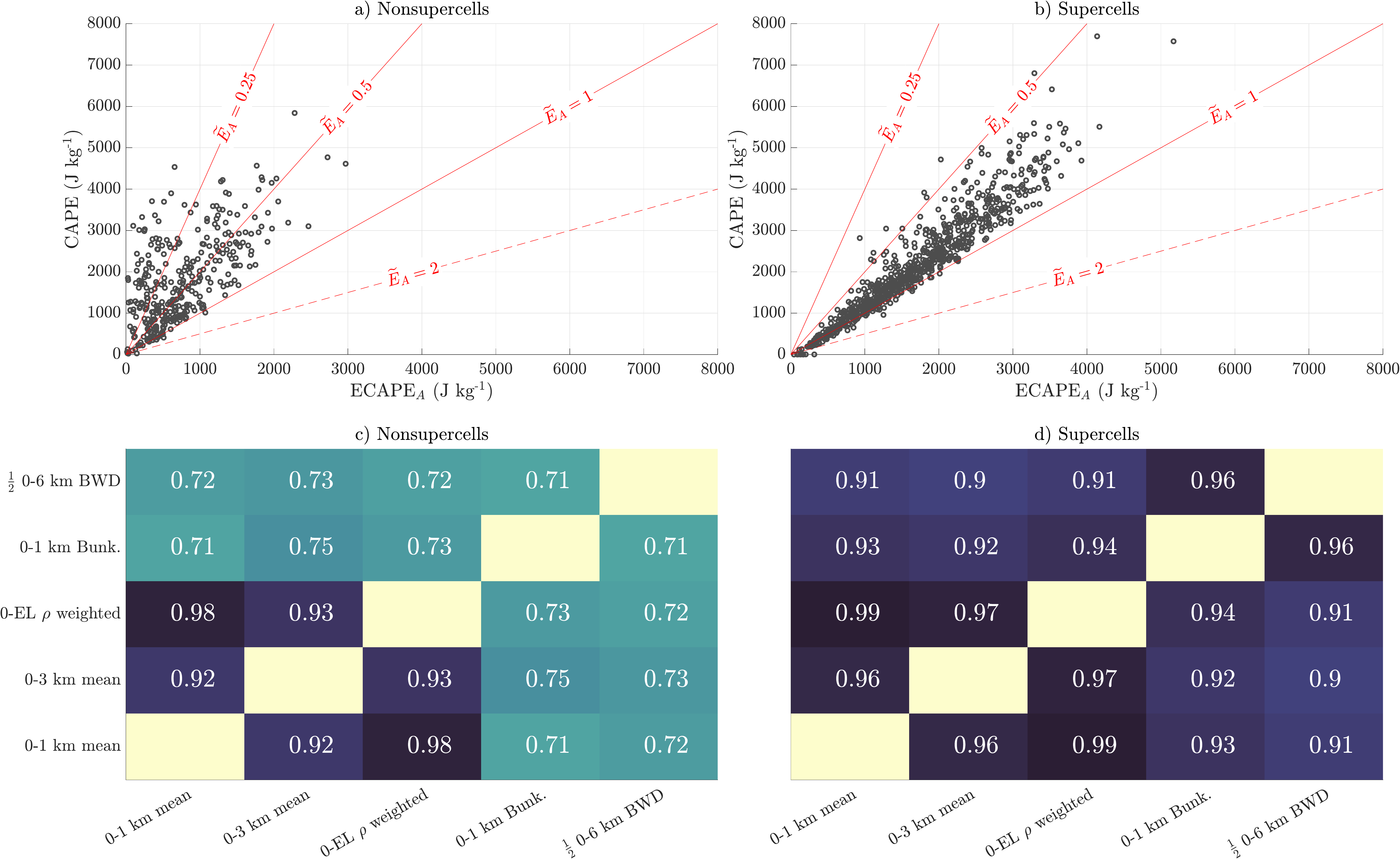}}
\caption{Top panels: scatter plots of ECAPE$_{A}$ ($x$ axis, J kg\textsuperscript{-1}) versus CAPE ($y$ axis, J kg\textsuperscript{-1}), computed with the \citet{THOM2003} soundings.  Panel a: 351 nonsupercell events, and panel b: 834 supercell events.  Contours of $\widetilde{E}_A$ are shown in red.  Panels c-d: $R^2$ between solutions for ECAPE$_{A}$ computed using different definitions of $V_{SR}$.  A given cell shows the correlation coefficient between ECAPE$_{A}$ computed with the $V_{SR}$ definition on the $x$ axis, with that on the corresponding $y$ axis, with colors corresponding to the relative magnitudes.  } \label{lastfig}
\end{figure*}

Our final analysis examines the distribution of ECAPE$_A$ within the \citet{THOM2003} sounding dataset.  Once again, we use the 0-1 km mean $V_{SR}$ computed with the observed storm motion in our formulas, though we evaluate other definitions of $V_{SR}$ later in this sub-section.  The distribution of ECAPE$_A$ for all nonsupercell severe weather events is plotted against undiluted CAPE in Fig. \ref{lastfig}a.  Contours of $\widetilde{E}_A$ (the fraction of CAPE ``realized") are also shown for reference.  There is substantial variability $\widetilde{E}_A$, with ECAPE$_A\approx$ CAPE ($\widetilde{E}_A\approx 1$) in some events, and ECAPE$_A<<$ CAPE ($\widetilde{E}_A<< 1$) in others.  Furthermore, case-to-case variations in ECAPE$_A$ and CAPE only loosely corresponded with one another, with $R^2=0.46$ based on a linear fit of these two quantities. In most events, particularly those with significant CAPE ($>1000 \; J/kg$), ECAPE$_A$ was less than CAPE suggesting that most nonsupercell storms only realize a fraction of their available CAPE.  %The poor correspondence between ECAPE and CAPE also suggests that CAPE is a relatively poor predictor of updraft buoyancy in nonsupercell thunderstorms.

In contrast with nonsupercell events, there is a much closer correspondence between ECAPE$_A$ and CAPE in supercell events, with $R^2=0.90$ between these two variables (Fig. \ref{lastfig}b).  Furthermore, $\widetilde{E}_A>0.5$ for nearly every supercell sounding, and this quantity was close to 1 in many cases, and exceeded 1 in a handful of instances.  This corroborates the idea, proposed by \citet{PNM2019}, that supercells realize a larger percentage of their environmental CAPE than nonsupercells.  The primary reason for this difference is the larger vertical wind shear, and consequently storm-relative flow, in supercell environments relative to nonsupercell environments.  Hence, CAPE may be a better predictor of storm-to-storm variations in updraft intensity in supercells than it is in nonosupercells.  However, there is still substantial variability in the correspondence between ECAPE and CAPE, particular for larger CAPE values, which suggests that ECAPE provides added value over CAPE in supercell environments.

To evaluate the sensitivity of ECAPE to how $V_{SR}$ is calculated, we re-computed ECAPE$_A$ with the 0-3 km mean $V_{SR}$ with the observed storm motion, the density weighted average of $V_{SR}$ below the LFC with the observed storm motion, the 0-1 km mean $V_{SR}$ computed using the storm motion estimate of \citet{B2000} which includes components of storm motion driven by advection and propagation, and the advective storm motion only, estimated as half the 0-6 km bulk wind difference.  Results with the $V_{SR}$ measures that use the observed storm motion yield nearly identical results to one another in both nonsupercells (Fig. \ref{lastfig}c) and supercells (Fig. \ref{lastfig}d), with $R^2$ ranging from 0.96 to 0.99.  

In the case of supercells, the ECAPE$_A$ computed with the observed storm motion corresponded well with the ECAPE$_A$ computed using the Bunkers storm motion estimate and half the bulk wind difference (Fig. \ref{lastfig}d).  However, this correspondence was degraded slightly in nonsupercell events, with the $R^2$ ranging form 0.71 to 0.75 between ECAPE$_A$ computed with the observed storm-motion, with that computed using the bunkers estimate and bulk wind difference.  This likely reflects the fact that the motion of nonsupercell storms is more often influenced by extraneous factors like outflow and airmass boundaries, than in supercells.  Hence, sounding-based estimates for storm motion do not correspond with actual storm motions as well in nonsupercell events as they do in supercell events.  

In many contexts where this formula would be used, such as in forecasting, the storm motion is unknown and must be estimated.  This analysis suggests that estimating storm motion with the method of \citet{B2000} or half the 0-6 km BWD are both viable choices.

\section{Summary, conclusions, and discussion}

In summary, we have derived a formula for ECAPE that depends entirely on state variables available within an atmospheric sounding.  This formula relies on three concepts: a scaling between fractional entrainment and updraft radius of $\varepsilon \sim R^{-2}$, the adiabatic conservation of moist static energy, and a direct correspondence between the cloud relative flow and the updraft radius.  Finally, we have accounted for the potential enhancement of updraft kinetic energy via pressure accelerations.  We recommend using the  following steps to compute this quantity in a software routine:
\begin{enumerate}
    \item 
    Set the following constant values: $c_{p}=1005$ J kg\textsuperscript{-1} K\textsuperscript{-1}, $L_{v,r}=$ 2,501,000 J kg\textsuperscript{-1}, $g=9.81$ m s\textsuperscript{-1}, $\sigma=1.6$, $\alpha=0.8$, $k^2 = 0.18$, $P_r=\frac{1}{3}$, and $L_{mix}=120$ m.
    \item
    Compute CAPE, the $LFC$, and the $EL$ for an undiluted parcel from an atmospheric profile using an existing software routine (e.g., SHARPy, Metpy).
    \item
    Compute the following parameter:
    \begin{equation}
    \label{eq:F2}
    \psi = \frac{k^2 \alpha^2 \pi^2 L_{mix}}{P_r \sigma^2 H},
    \end{equation}
    where $H$ is the equilibrium level.
    %%,
    \item
    Compute $V_{SR}$ from an atmospheric profile.  We recommend averaging $V_{SR}$ in the 0-1 km layer, using the method for estimating storm motion described by \citet{B2000}.
    \item
    Evaluate the following formula, using a numerical integration scheme.
    \begin{equation}
    \label{eq:F4}
        \widehat{h_0}(z) = \frac{1}{z} \int_{z^*=0}^{z^*=z} \left( c_{pd} T_0 + L_{v,r} q_0 + gz^* \right) dz^*,
    \end{equation}
    This procedure only needs to be done once in a given profile, and yields $<h_0>$ as a function of height.
    \item
    Compute NCAPE, using the following formula:
    \begin{equation}
    \label{eq:F3}
    \text{NCAPE} = - \int_{z=LFC}^{z=EL} \frac{g}{c_{pd} T_0} \left( \widehat{h_0} -  h_0^* \right) dz,
    \end{equation}
    NCAPE is positive in most contexts though it may become negative in environments with large free tropospheric relative humidity.
    \item
    Compute ECAPE$_A$, using the following formula:
    %%$
    %%$
    \small
    \begin{equation}
    \label{eq:F1}
    \begin{aligned}
    \text{ECAPE}_A = \frac{V_{SR}^2}{2} + \frac{-1 - \psi - \frac{2 \psi}{V_{SR}^2} \text{NCAPE}}{4 \frac{\psi}{V_{SR}^2}} + \\ \frac{\sqrt{\left(1 + \psi + \frac{2\psi}{V_{SR}^2} NCAPE \right)^2 + 8\frac{\psi}{V_{SR}^2} \left( \text{CAPE} - \psi \text{NCAPE}\right) }}{4 \frac{\psi}{V_{SR}^2}}.
    \end{aligned}
    \end{equation}
    \normalsize
    In the case of a negative solution to this equation, set the $\text{ECAPE}_A$ to 0. 
\end{enumerate}

Our results show that ECAPE provides a more accurate prediction of updraft intensity than standard CAPE when forecasting severe weather hazards that depend on middle-to-upper tropospheric vertical velocities.  Examples of these situations include forecasting heavy precipitation, large hail, and intense cold pools and downdrafts.  Hence, it would benefit the forecasting community to display this quantity alongside standard CAPE on websites that provide numerical weather prediction model output graphics, such as the storm-prediction center Mesoanalysis site.  In addition, $\widetilde{E}_A$, which is the fraction of CAPE realized, is a powerful discriminator of supercellular from nonsupercellular storm mode, with a True Skill Statistic \cite[TSS; e.g., section 2 in][]{PNM2020b} of 0.76 in this prediction.  This is on par with the TSS for 0-1 km $V_{SR}$, which is 0.79 (these values are not statistically different).  The physical reason behind this discriminatory skill relates to the conclusions of \citet{PNM2019}, who showed that supercells realize larger fractions of their CAPE than nonsupercells (and hence have larger $\widetilde{E}_A$).

A variety of research applications would also benefit from the consideration of ECAPE, in addition to standard CAPE.  For instance, studies in past literature often contrast storm dynamics in high-shear low-CAPE severe weather events with events \citep[e.g.,][]{SD2008} occurring in environments with higher CAPE (and sometimes weaker shear).  The premise behind this distinction is, because of the small updraft buoyancy in low-CAPE events, the updrafts accelerations in these storms are dominated by dynamic pressure accelerations rather than buoyancy \citep{WP2021}.  However, it is possible that because of the extreme shear in many low-CAPE severe weather outbreaks, updrafts in these scenarios realize a higher percentage of their CAPE than their counterparts in high CAPE environments.  Hence, ECAPE may more accurately distinguish between storms with large and small buoyancy than standard CAPE, and a reconsideration of the analyses in these past studies with distinctions drawn between high ECAPE and low ECAPE events may yield additional insights into storm dynamics. 

ECAPE may also yield novel insight into the influence of climate change on thunderstorms.  For instance, a subset of studies that investigate the influence of climate change on severe storm behavior use proxy analyses in global climate model (GCM) simulations, assessing the impacts of global warming on parameters like CAPE and CIN.  Future changes to free tropospheric relative humidity, temperature, and vertical wind shear are also likely to influence thunderstorms via the connection between these environmental attributes and entrainment.  Investigating changes to the climatology of ECAPE in future climates is a concise way of encapsulating these yet-to-be explored climate change influences on storm entrainment, and consequently storm intensity.  Efforts to quantify the effects of climate change among the authors of the present study are currently underway.

Some of the intermediary formulas that express buoyancy and ECAPE as an analytic function of fractional entrainment may be useful in cumulus parameterization schemes.  For instance, multi-plume schemes like the scheme of \citet{AS1974}, the Relaxed Arakawa-Schubert scheme \citet{RAS1992}, the EDMF$^N$ scheme \citet{RN2015}, and the MAP scheme \citep{PMZ2020} require the computation of diluted buoyancy and ECAPE for each plume.  In the traditional approach for computing ECAPE, these schemes would execute two numerical vertical integrations for each plume.  This procedure, however, is dramatically simplified by using eq. \ref{eq:ECAPE_analytic2} in the present study, where only 3 vertical integrations per grid cell are needed to obtain CAPE and NCAPE, and then the ECAPE associated with each plume is computed analytically.  The MAP scheme from \citep{PMZ2020} was also formulated to use the formula from P20 as part of its closure for convective mass flux.  The formula presented here is a more accurate alternative.

A potential caveat to using this parameter operationally is that ECAPE$_A$ vanishes in the absence of $V_{SR}$, whereas we know that deep convection is possible in the absence of substantial $V_{SR}$.  This discrepancy is likely a consequence of the primary controls on updraft width shifting away from vertical wind shear to other environmental factors when shear is weak, such as the planetary boundary layer (PBL) depth \citep[e.g.,][]{MPM2021b} or the width scale of terrain features \citep[e.g.,][]{NMP2021,KIR2022}.  A potential way to circumvent this issue is to revert to a standard ECAPE calculation (with a user-prescribed $\varepsilon$) in these weakly sheared environments, setting the updraft radius to scale with the PBL depth or to a constant value \citep[e.g., 1500 m, as was done in][]{PMZ2020}.

Some may debate the semantics over whether the formulas derived are more appropriately described as predictive equations for the maximum updraft vertical velocity, rather than a modified CAPE that accounts for entrainment.  Some view CAPE as pertaining only to an isolated ascending parcel with no explicit assumptions about updraft structure and behavior.  Hence, our inclusions of updraft dynamics in our ECAPE calculation makes this calculation conceptually distinct from that of CAPE.  However, we argue that there are a variety of conceptual definitions of CAPE in past literature, and that this quantity is often used in the forecasting community to predict how a given thermodynamic environment may affect updraft vertical velocity.  Because of the familiarity of forecasters with CAPE, ECAPE (with units of J kg\textsuperscript{-1}) is a more relatable quantity to forecasters than $w_{max}$.  This is the primary reason why we have advertised the quantity derived here as an ECAPE, rather than a predictor of $w_{max}$.

\clearpage
%%%%%%%%%%%%%%%%%%%%%%%%%%%%%%%%%%%%%%%%%%%%%%%%%%%%%%%%%%%%%%%%%%%%%
% ACKNOWLEDGMENTS
%%%%%%%%%%%%%%%%%%%%%%%%%%%%%%%%%%%%%%%%%%%%%%%%%%%%%%%%%%%%%%%%%%%%%
\acknowledgments
J. Peters's efforts were supported by National Science Foundation (NSF) grants AGS-1928666, AGS-1841674, and the Department of Energy Atmospheric System Research (DOE ASR) grants DE-SC0000246356. D. Chavas was supported by National Science Foundation (NSF) grants 1648681 and 2209052. H. Morrison was supported by DOE ASR grant DE-SC0020104. The National Center for Atmospheric Research is sponsored by NSF.  

%%%%%%%%%%%%%%%%%%%%%%%%%%%%%%%%%%%%%%%%%%%%%%%%%%%%%%%%%%%%%%%%%%%%%
% DATA AVAILABILITY STATEMENT
%%%%%%%%%%%%%%%%%%%%%%%%%%%%%%%%%%%%%%%%%%%%%%%%%%%%%%%%%%%%%%%%%%%%%
% 
%
\datastatement
Matlab code to compute ECAPE using an atmospheric sounding as input is available at https://doi.org/10.6084/m9.figshare.21859818.
\bibliographystyle{ametsocV6}
\bibliography{references}

\end{document}